\documentclass[twocolumn]{emulateapj}
\shorttitle{ High Reliability of {\it Kepler}'s Multiple Planet Candidates }
\shortauthors{ Lissauer et al.}
\usepackage{natbib}
\newcommand{\ikt}{{\it Kepler}}
\newcommand{\ik}{{\it Kepler~}}

\begin{document}
\slugcomment{ApJ in press}
\title{ Almost All of \ikt's Multiple Planet Candidates are Planets}

\author{Jack J. Lissauer\altaffilmark{1}, Geoffrey W. Marcy\altaffilmark{2}, Jason F. Rowe\altaffilmark{1,3}, Stephen T. Bryson\altaffilmark{1}, Elisabeth Adams\altaffilmark{4}, Lars A. Buchhave\altaffilmark{5,6}, David R. Ciardi\altaffilmark{7}, William D. Cochran\altaffilmark{8}, Daniel C. Fabrycky\altaffilmark{9,10}, Eric B. Ford\altaffilmark{12},  Francois Fressin\altaffilmark{4}, John Geary\altaffilmark{4}, Ronald L. Gilliland\altaffilmark{13}, Matthew J. Holman\altaffilmark{4}, Steve B. Howell\altaffilmark{1}, Jon M. Jenkins\altaffilmark{1,3}, Karen Kinemuchi\altaffilmark{1,14}, David G. Koch\altaffilmark{1}, Robert C. Morehead\altaffilmark{12}, Darin Ragozzine\altaffilmark{4}, Shawn E. Seader\altaffilmark{1,3}, Peter G. Tanenbaum\altaffilmark{1,3},
Guillermo Torres\altaffilmark{4}, Joseph D. Twicken\altaffilmark{1,3}
}

\email{ Jack.Lissauer@nasa.gov }
\altaffiltext{1}{NASA Ames Research Center, Moffett Field, CA 94035, USA}
\altaffiltext{2}{Astronomy Department, University of California, Berkeley, CA 94720, USA}
\altaffiltext{3}{SETI Institute/NASA Ames Research Center, Moffett Field, CA 94035, USA}
\altaffiltext{4}{Harvard-Smithsonian Center for Astrophysics, 60 Garden Street, Cambridge, MA 02138, 
USA}
\altaffiltext{5}{Niels Bohr Institute, University of Copenhagen, DK-2100, Copenhagen, Denmark} 
\altaffiltext{6}{Centre for Star and Planet Formation, Natural History Museum of Denmark, University of Copenhagen, DK-1350 Copenhagen, Denmark}
\altaffiltext{7}{Exoplanet Science Institute/Caltech, Pasadena, CA 91125, USA}
\altaffiltext{8}{Department of Astronomy, University of Texas, Austin, TX 78712, USA}
\altaffiltext{9}{Department of Astronomy \& Astrophysics, University of California, Santa Cruz, CA 95064, 
USA}
\altaffiltext{11}{Hubble Fellow}
\altaffiltext{12}{University of Florida, 211 Bryant Space Science Center, Gainesville, FL 32611, USA}
\altaffiltext{13}{Space Telescope Science Institute, Baltimore, MD 21218, USA}
\altaffiltext{14}{Bay Area Environmental Institute, CA}

\begin{abstract}
We present a statistical analysis that demonstrates that the overwhelming majority of \ik candidate 
multiple transiting systems (multis) indeed represent true, physically-associated transiting planets.  Binary stars provide the primary source of false positives among \ik planet candidates, implying that false positives should be nearly randomly-distributed among \ik targets.  In contrast, true transiting planets would appear clustered around a smaller number of \ik targets if detectable planets tend to come in systems and/or if the orbital planes of planets encircling the same star are correlated.  There are more than one hundred times as many \ik planet candidates in multi-candidate systems as would be predicted from a random distribution of candidates, implying that the vast majority are true planets.  Most of these multis are  
multiple planet systems orbiting the \ik target star, but there are likely cases where (a) the planetary
system orbits a fainter star, and the planets are thus significantly larger than has been estimated, or (b)
the planets orbit different stars within a binary/multiple star system.
We use the low overall false positive rate among \ik multis, together with analysis of \ik spacecraft and ground-based data, to validate 
the closely-packed Kepler-33 planetary system, which orbits a star that has evolved somewhat off of the 
main sequence.  Kepler-33 hosts five transiting planets with periods ranging from 5.67 to 41 days.

\end{abstract}

\clearpage

\section{ Introduction }

Roughly one-third of \ikt's planet candidates announced by \cite{Borucki:2011} are associated with 
targets that have more than one candidate planet.  False positives (FPs) plague ground-based transit 
searches, but the exquisite quality of \ik photometry, combined with the ability to measure small 
deviations in center of light during transits \citep{Jenkins:2010, Batalha:2010}, have been used to 
cleanse the sample prior to presentation in \cite{Borucki:2011}. Accounting for candidates on each one's individual merit, \cite{Morton:2011} estimated the fidelity of \ikt's planet 
candidates (fraction of the candidates expected to be actual planets) to be above $90\%$.  Yet the fidelity of 
multiple planet candidates is likely to be higher than that for singles \citep{Latham:2011,              Lissauer:2011a}.  We show herein that the vast majority of \ikt 's multiple planet candidates are true multiple planet systems.

The majority of \ik planet discoveries announced to date have been confirmed dynamically, using radial 
velocity variations of the target star or departures of observed transit times from a linear ephemeris 
(transit timing variations, TTVs).  However, three of the first sixteen \ik planets do not have dynamical 
evidence supporting the discovery, but rather these planets have been ``validated'' by showing that the 
probability that the observed signal is produced by a planet around the host star is at least 100 times as 
large as that of a false positive \citep{Torres:2011, Fressin:2011, Lissauer:2011b}.  For these three planets, the validation process was based on the
results of the  {\tt BLENDER} code, which performs detailed comparisons
between the observed \ik lightcurve and the lightcurves predicted
for an ensemble of theoretically conceivable FPs to reject FP
models that are not consistent with the observations.   {\tt BLENDER} then
sums  the {\it a priori} probability for the allowed false positives
and compares that to the {\it a priori} probability for the planet
model.   {\tt BLENDER} validates a planet when the resulting odds ratio
very strongly favors the planet model.
Although all 
three of these planets orbit stars that also have dynamically-confirmed planets, planetary multiplicity was 
not used quantitatively to aid in their validation.

\cite{Morton:2011} do a more cursory analysis, but one that is far easier, enabling them to perform it  for 
all of the candidate planets (``candidates'', henceforth)  that are listed in the \cite{Borucki:2011} paper.  The 
study  by \cite{Morton:2011} assumes a flat 20\% planet occurrence rate as a prior for their Bayesian analysis (i.e., the average number of planets per star is assumed to be 0.2), and they do not consider planets orbiting background stars to be FPs.  \cite{Morton:2011} use  the depth and period of the transit, together with the magnitude, 
stellar properties and galactic latitude of the target star (henceforth ``target'') to provide an upper bound 
on the FP prior; for high SNR transits, this upper bound on the FP prior can be much higher than the one 
computed using detailed lightcurve matching by {\tt BLENDER} to rule out many FP scenarios \citep
{Torres:2011, Fressin:2011}. Indeed, \cite{Morton:2011}  do not examine \ik lightcurves, but rather 
assume for their calculations that  various false-positive vetting procedures using \ik photometry have been done, such as 
elimination of candidates with  V-shaped lightcurves or where the centroid of light received on the \ik 
focal plane is in a different position during a transit than it is outside of transit, and that the planet candidates show 
no evidence for being FPs.  However, these cuts were not applied to all of the candidates listed in 
\cite{Borucki:2011}.  Some candidates were not vetted as extensively as assumed by \cite{Morton:2011}, and other 
candidates did not pass one of the cleanness tests but still remain viable, if somewhat more suspect, 
planetary candidates (e.g., grazing planetary transits, as well as grazing stellar eclipses, produce V-shaped 
lightcurves), and thus remain on \cite{Borucki:2011}'s candidate list. 

We develop herein a quantitative method to estimate the true planet fraction of the overall sample of  
\ik candidate multiple planets that is based upon the observation that there are far more multiples than 
would be expected for a random distribution of candidates among \ik targets \citep{Lissauer:2011a}. This 
``overabundance'' of multiples is illustrated by the fact that fewer than 1\% of targets have a candidate on \cite{Borucki:2011}'s candidate list, 
whereas more than 15\% of targets with at least one candidate have multiple candidates, and more than 
25\% of targets with at least two candidates have a third candidate. Such a result is expected if planets 
often appear within systems
of planets, especially if such systems tend to be flat in the distribution of
inclinations. High multiplicity, low inclination, planetary systems are consistent with \ik observations \citep{Lissauer:2011a}.

We describe an approach to validation of specific multiple planet system candidates that is intermediate 
between those of \cite{Morton:2011} and {\tt BLENDER}, and we apply these techniques to validate the very flat and dynamically-packed KOI-707 system\footnote{Targets with planetary candidates are given an integer KOI (\ik Object of Interest) number, and individual candidates are denoted by the target KOI number followed by a decimal point and a two digit number specifying the order in which they were identified.   For example, KOI-707.02 is the second planetary candidate to have been identified by transit-like signatures in the lightcurve of the target star KOI-707.}. In this approach, applied herein exclusively to KOI-707, we examine the target and 
candidates individually, supplementing the \ik Input Catalog \citep[KIC,][]{Brown:2011} with spectroscopic 
observations of the target and high-resolution imaging, including Keck guider camera exposures and 
speckle imaging \citep{Howell:2011} and/or adaptive optics \citep{Troy:2000, Hayward:2001}, to search 
for other stars in the target aperture.  We check the lightcurve for a shape consistent with a transit. We 
verify that the centroid motion is nil or consistent with a transit around the target star and not a transit/
eclipse of any other stars observed nearby. But we do not perform detailed matching of observed 
lightcurves to FP scenarios. 
We then use multiplicity to effectively increase the planet priors and allow for validation in a Bayesian 
sense.  Note that for candidate giant ($\sim$ 1 Jupiter radius) planets, one would need to consider the possibility of the transit signal being produced by a late M star partially eclipsing the target star (perhaps 
with dilution); for smaller candidates, the possibility of transits of fainter stars 
by larger planets must be taken into account.  

The probability of a  candidate in a multiple system being a false positive is contrasted to that of a 
single candidate in Section 2.  Other factors involved in validating transiting planet candidates are 
discussed in Section 3.  We apply our method to the KOI-707 candidate multi-planet system in Section 4.  
We conclude in Section 5 with a summary of our results and a discussion of their implications.

\section{ Statistical Validation of Multi-Planet Candidates }

\ik has found far more multiple candidate systems than would be the case if candidates were randomly 
distributed among target stars \citep{Lissauer:2011a}. False positives are expected to be nearly 
randomly distributed among \ik targets, whereas true transiting planets could be clustered if planets 
whose sizes and periods are adequate for transits to be detected often come in multiples, as is the case 
for planets detected by radial velocity variations \citep{Wright:2009}, and/or if planetary systems 
tend to be flat, so geometry leads to higher transit probabilities for other planets if one planet is seen to 
transit  \citep{Ragozzine:2010}. We quantify the excess of multiples and its implications for the overall 
fidelity of the sample of multiple candidates in this section.

 We use the following notation and input parameters: The number of planet candidates is $n_c = 1199$, 
the number of targets with two or more candidates, i.e., the number of candidate multiple planet systems, 
is $n_m = 170$, and the number of targets from which the sample is drawn is $n_t$ = 160,171.  These 
numerical values are taken from \cite{Lissauer:2011a}, who removed from the 1235 candidates listed in 
\cite{Borucki:2011} those objects that were estimated to have radii greater than twice that of Jupiter, 
those for which only one transit had been observed, and those identified as FPs by \cite{Howard:2011}. 
We denote the number of actual planets among the candidates as $n_p$, the number of false positives 
as $n_f$, and the fidelity of the sample (fraction of candidates that are planets) as $P$, so $n_c = n_p + 
n_f$ and $P = n_p/n_c$.  

For our calculations in Equations (1--9), we make the following two assumptions:

\noindent i) FPs are randomly distributed among the targets; 

\noindent ii) there is no correlation between the probability of a target to host one or more detectable 
planets and to display FPs.

\noindent In contrast, we do not assume that planets are randomly distributed among the targets. The previous verifications of the two-planet
Kepler-10 system \citep{Batalha:2011, Fressin:2011}, the three-planet
Kepler-9 \citep{Holman:2010, Torres:2011} and Kepler-18 systems
\citep{Cochran:2011}, and especially the six-planet Kepler-11 system
\citep{Lissauer:2011b}, provide strong evidence of the non-random distribution of transiting planets among \ik target stars. 

The above two assumptions, together with observed values of $n_c$, $n_m$ and $n_t$ and an 
assumed lower bound on $P$, allow us to estimate lower bounds on the fidelity of candidates in various 
classes of candidate multi-planet systems using simple algebra.  For example, assumption (1) implies 
that the expected number of targets with $k$ false positives, $E(k)$, is given by a Poisson distribution of $n_t$  
members whose mean is given by the average number of false positives per target, $\lambda = (1-P)
n_c/n_t$.  This expectation value is given by the following formula: 

\begin{eqnarray}
 E(k) = \frac{\lambda^k e^{-\lambda }}{k!} n_t &=& \frac{((1-P)\frac{n_c}{n_t})^k e^{-(1-P)\frac{n_c}{n_t}}}{k!}
n_t \\
&\approx& \frac{((1-P)\frac{n_c}{n_t})^k}{k!}n_t , \nonumber
\end{eqnarray}
 
\noindent where the approximation in Equation (1) is valid for $\lambda \ll 1$. 

We next compute the expected number of \ik targets with multiple candidates at least one of which is a 
FP.  In some cases, we make approximations that yield conservative estimates of the expected numbers 
of true planets (i.e., overestimate the expected number of FPs).
Our results, presented in Equations (2--7), are given first as general formulae, then as a function of the 
fraction  of candidates that are true planets, $P$, for the observed values of 160,171 targets, 1199  
candidates, and 170 multiple candidate systems stated above, and finally these value if $P$ = 0.5 or 0.9\footnote{We choose to present numerical results for $P = 0.9$ because we consider it to be a reasonable estimate of the actual planet fidelity rate, and $P = 0.5$ to show that even for an overall FP rate far higher than studies suggest, the expected number of FPs among the multis is remarkably low.}.  These final numbers are to be 
compared to the observation that there are 115 candidate 2-planet systems, with a total of 230 planet 
candidates, and 55 candidate systems of 3 or more transiting planets, with a total of 178 planet 
candidates. 

Equation (1) yields estimates of the number of targets with two or three false positives:

\begin{eqnarray}
2 ~{\rm FPs:} ~ \frac{((1-P)n_c)^2}{2n_t}&=& 4.49 (1-P)^2 \\
  &=&  1.12 ~{\rm or}~ 0.045; \nonumber
\end{eqnarray}
\begin{eqnarray}
 3 ~{\rm FPs:} ~\frac{((1-P)n_c)^3}{6n_t^2} &=&  0.011 (1-P)^3   \\
 &=&  0.0014 ~{\rm or}~ 1.1 \times 10^{-5}.\nonumber
\end{eqnarray}

\noindent The number of targets expected to have four or more false positives is very small and can be 
neglected for our purposes. Equation (1) also yields the expected number of targets with a single FP under the same assumptions, 597.3 or 119.8.

The assumed lack of correlation between the propensity of a target to have false positives and true 
planets implies that the probability that a given target hosts both a planet and one or more false positive 
is equal to the product of these individual probabilities.  The probability
of a target having a detected transiting planet is given by $Pn_c/n_t$, and the probability of it showing a false positive
is given by
Equation (1).  Thus, the estimated numbers of targets with at least one planet as well as one or more false positives 
are given by:
  
\begin{eqnarray}
1 ~{\rm planet}~ + 1~{\rm FP:}&~& \frac{n_cP}{n_t}\times \frac{n_c(1-P)}{n_t} \times n_t \\
&=& P(1-P) \frac{n_c^2} {n_t}\nonumber \\
 &=& 8.98 P(1-P) = 2.25 ~{\rm or}~ 0.81; \nonumber
\end{eqnarray}
\begin{eqnarray}
1 ~{\rm planet}~ + 2~{\rm FPs:}&~& P(1-P)^2 \frac{n_c^3}{2n_t^2} \\
&=& 0.034 P(1-P)^2 \nonumber \\
&=& 0.0042 ~{\rm or}~ 3 \times 10^{-4}.  \nonumber
\end{eqnarray}

\noindent The number of targets expected to have both a planet and three or more false positives is very 
small and can be neglected for our purposes. Somewhat lower estimates of the number of multi-
candidate systems with one planet and at least one FP than given by Equations (4) and (5) can be 
derived by noting that the observed multiplicity implies fewer (than $n_c$) targets with a nonzero 
number of planet candidates.

To derive an estimate of the expected number of targets with multiple true planets as well as at least one 
FP, replace the term in Equations (4) and (5) that represents the probability of a given target having a 
true planet, $Pn_c/n_t$, with a term representing the probability of a given target having a multi-planet 
system, $n_m/n_t$. (Assuming all multis are multi-planet systems to estimate of this term is conservative in the sense that it may lead to an overestimate in the 
number of false positives in multi-candidate systems.)

 \begin{eqnarray}
{\rm 2~or~more~planets}~ + 1~{\rm FP:}&~& \frac{n_m}{n_t}\times \frac{n_c(1-P)}{n_t} \times n_t \\
&=& (1-P)\frac {n_mn_c}{n_t} = 1.27 (1-P)  \nonumber\\
 &=& 0.64 ~{\rm or}~ 0.13; \nonumber
\end{eqnarray}

\begin{eqnarray}
{\rm 2~or~more~planets}~ + 2~{\rm FPs:}&& \frac{n_m}{n_t}\times(1-P)^2 \times \frac{n_c^2}{2n_t} \\
&=& 0.0048 (1-P)^2 \nonumber \\
&=& 0.0012 ~{\rm or}~ 5 \times 10^{-5}. \nonumber
\end{eqnarray}

\noindent The number of targets expected to have both multiple planets and three or more false 
positives is extremely small and can be neglected for our purposes. 

\subsection{Numerical Estimates for \ik Candidates}

Adding up the above estimates of FPs of different types yields expected totals of 4.5 or 0.9 FPs (for assumed FP rates of 50\% and 10\%, respectively) in 
doubles (compared to a total of 230 candidates) and 0.65 or 0.13 FPs in systems with three or more 
candidates (compared to a total of 178 candidates). These numbers suggest that validation of doubles at 
the 99\% level by this method alone is marginal. Triples and higher multiplicity candidates are more 
strongly validatable, enough so that the one dynamically-unstable candidate triple system noted by \cite
{Lissauer:2011a}, KOI-284, which has a pair of candidates with a period ratio $1.0383$, should be 
viewed as somewhat surprising (but see Section 2.4).  

It is of interest to compare the (low) false positive rates for multis 
estimated in the previous paragraph with the presumed FP rates for 
singles upon which their calculations are based. For 50\% FPs in 
singles, our estimates above correspond to 2\% FPs in doubles (compare the first numbers at the end of equations 2 and 4 with the 230 candidates in \ik candidate 2-planet systems) and  0.4\% FPs for targets with 
three or more candidates (compare the first numbers at the end equations 3 and 5--7 with the 178 candidates in \ik candidate 3 or more planet systems); for 10\% FPs in singles, we estimate 0.4\% FPs 
in doubles  (compare the final numbers at the end of equations 2 and 4 with the 230 candidates in \ik candidate 2-planet systems) and  0.1\% for targets with three or more candidates (compare the final numbers at the end equations 3 and 5--7 with the 178 candidates in \ik candidate 3 or more planet systems).  These 
values are consistent with a ``multiplicity boost'' in the Bayesian 
prior for planets of $\sim$ 25 for a system with one additional candidate 
and  of $\sim$ 100 for a system with more than one additional candidate.  
The probability that a target hosts a 
planet candidate, 1/150, can be compared to the probability of a target that hosts one 
candidate hosts another, 1/6, and that an additional candidate has been 
found for higher multiplicity systems, 1/3. These latter values, together with the assumption 
that FPs and true planets are uncorrelated, lead to an estimate 
of the multiplicity boost for candidate doubles of 30, slightly larger than the value of 25 computed above, and to a somewhat 
reduced 
estimate of $\sim$ 50 for candidates in systems of higher multiplicity.

The multiplicity boost introduced in the previous paragraph can be used to quantify the increased probability that a set of transit-like signatures represent a real transiting planet if one or more additional set of transit-like signatures are detected for the same \ik target; i.e., to compute an estimate the probability of a particular candidate in a multiple candidate system being a true planet from an estimate of that candidate being a planet that does not account for multiplicity.  We denote the probability of planethood of the candidate in question computed without factoring in multiplicity by $P_1$ and the probability of it being a planet accounting for multiplicity by $P_2$ if it is in a two candidate system and by $P_{3+}$ if it is a member of a three or more candidate system.  In our formulation, multiplicity effectively increases the  Bayesian estimate of the planet prior but leaves the FP prior unchanged.  Thus the probability of the planet interpretation accounting for multiplicity is related to that neglecting multiplicity according to the following approximate 
formulae:
 \begin{eqnarray}
P_2&\approx& \frac{25P_1}{25P_1+ (1-P_1)}\\
 &~(& \approx 1 - \frac{1-P_1}{25} ~{\rm for}~ 1-P_1 \ll 1 , ~{\rm and}~\approx 25P_1 ~{\rm for}~ P_1 \ll 1) \nonumber
\end{eqnarray}
\begin{eqnarray}
P_{3+} &\approx& \frac{50P_1}{50P_1+ (1-P_1)} \\
&(& \approx 1 - \frac{1-P_1}{50} ~{\rm for}~ 1-P_1 \ll 1 , ~{\rm and}~\approx 50P_1 ~{\rm for}~ P_1 \ll 1) , \nonumber
\end{eqnarray}
\noindent where in each case we use the smaller value computed in the previous paragraph. Rough estimates for the planet probabilities of individual \ik candidate multis can be obtained 
by applying Expressions (8) and (9) to the estimates of planet probability by \cite{Morton:2011}; the \cite
{Morton:2011} planet probabilities are given by $P_1$ in the above formulae since their numbers do not 
account for multiplicity.  Note that this multiplicity boost incorporates self-consistently both the coplanarity 
boost  discussed in \cite{Cochran:2011} and the systems boost (that planets tend to come in systems) 
without need to disentangle these effects.

The above calculations of the multiplicity boost do not assume that any of the candidates of the target under consideration have been confirmed or validated as planets. Does validating one or more candidate as a true planet(s) substantially increase the probability that other candidates of the same target are also planets?   For realistic estimates of FP rates (e.g., 10\%), the equations above show that most FPs in multis are likely to be combinations of 1 FP with one or more true planets, so validating one candidate does not substantially increase confidence in the others from a statistical viewpoint.  Nonetheless, validating one candidate does eliminate some of the non-statistical issues raised in the following three subsections. In particular, if that validated candidate is shown to not only be a planet but also to orbit the \ik target star, then the probability that the other candidates are physically-associated with the target increases.

\subsection{Caveats}

To examine the validity of the three assumptions stated above, we divide FPs into two classes, (I) 
chance-alignment blends such as background eclipsing binaries (BGEBs)  and (II) physical triple stars (PTs), where 
 grazing eclipses of the target star by a stellar companion are considered together with PTs for this purpose.  
We note that it is a matter of choice how to count cases where the transit signal is produced by a planet transiting a star other than the \ik target. Larger planets transiting stars that are not physically associated with the star providing most of the 
target's flux can be considered to fall within the chance-alignment blends/BGEB class of FPs (as they must 
be for \ikt's statistical 
census of planets, because background stars are not in the denominator of the
ratio taken to calculate the fraction of stars with planets, so their planets must not be included in the 
numerator), or alternatively not be regarded as FPs when assessing individual candidates and the 
reliability of the sample. Planets transiting physical companions to the star providing most of the flux can 
either be viewed as members of the PT class of FPs or not classified as FPs; how these planets are 
categorized for  \ikt's statistical 
census of planets is also a matter of choice.  

 Assumptions (i) and (ii) lead to the
following caveats:

(i): The expected number of BGEBs (chance-alignment blends) that can cause FPs varies with the 
magnitude and galactic latitude of the target star, whereas those of PTs do not (except to the extent that 
the distribution of stellar types among \ik targets depends somewhat on magnitude).  The variations in 
expected BGEBs act to increase the expected total number of systems with 2 or more FPs compared to the 
value given by the above formulae.  When looking at individual candidate systems, the FP probability of 
those around faint targets and at low galactic latitude is larger, whereas the planetary interpretation 
becomes more likely for brighter targets and those at higher galactic latitude.  However, the distributions in galactic latitude of the
planet candidates and of the multiple candidate systems both track
that of all \ik targets, whereas the distribution of expected BGEB FPs
is quite different (Figure 1); this suggests that
the fraction of candidates (both singles and multis) that are produced by BGEBs is small.

%\epsscale{0.9}
\begin{figure}
\includegraphics[scale=0.63,angle=0]{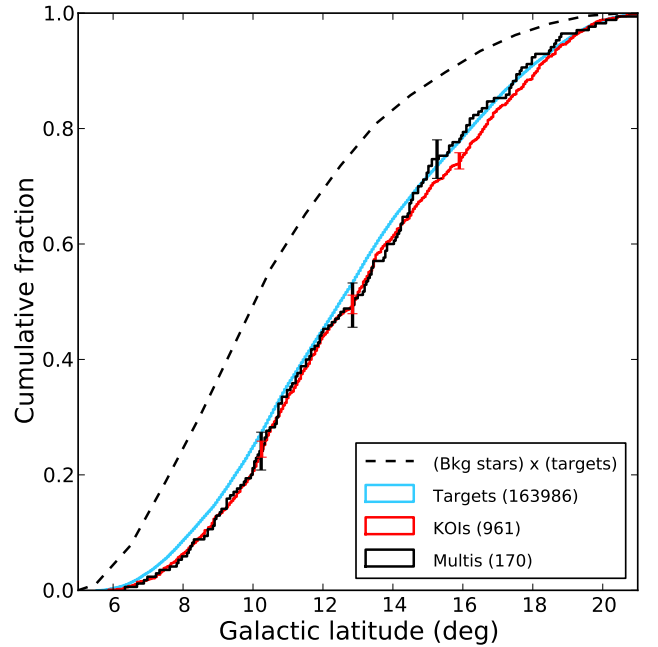}%{galacticlat.eps}   F1
\caption{Distributions in galactic latitude of all 163,986 \ik Quarter
2 exoplanet targets (solid turquoise curve), the 961 targets with planet
candidates meeting our criteria (solid dark red curve) and the 170 multiple
candidate systems (solid black curve). The dashed curve shows the
distribution of expected BGEB FPs, which we compute by weighting each
target by the sky density of stars with magnitudes between 15.0 and
20.0 according to the Besancon model of the galaxy (Robin et al.~2003,
2004).  All curves are normalized to give a total of unity.  Error bars represent the statistical uncertainty 
(1$\sigma$ binomial variation) of each of the planet candidate curves at quartile points in their 
distributions. The basic
shape of the three solid curves is governed by the orientation of the
\ik field on the sky, where the amount of area covered increases from
a point at galactic latitude $5^\circ$ to a region spanning $16^\circ$ in
longitude at $13^\circ$ in galactic latitude and then decreases to a
point at $21^\circ$ latitude; the second most important factor is a
weighting towards the galactic plane because of the higher spatial
density of target stars in that portion of the field.  The dashed
curve is more heavily weighted towards the galactic plane because, to
first approximation, it is weighted quadratically in the spatial
density of stars (more precisely, the product of the spatial densities
of target stars and background stars), whereas the other curves are
only weighted linearly.  The spatial density of background stars drops
by a factor of 17 from the lowest latitude \ik observes to the highest
latitude. The similarity of curves for targets, planets and multis,
and the difference of these from the curve estimating BGEB FPs, suggests
that the fraction of candidates that are produced by BGEBs (as well as
that resulting from planets transiting background stars) is small. \vspace{0.4 in}
}
\end{figure}

The statistical 
analysis that we presented above does not discriminate between multiple 
candidates resulting from multiple planets orbiting a background star and a single 
candidate resulting from a single planet orbiting a background star.  However, the correlations and lack 
thereof between the curves shown in Figure 1 also imply that the fraction of KOIs caused by single or 
multiple planets transiting a background star is small. In some cases, it is especially unlikely that a multiple
candidate system is produced by several planets orbiting a background
star, either because its largest member would be greater than
planetary in size, or because the requirement of dynamical stability
would imply implausibly small densities for the larger planets
orbiting the fainter star.  Additionally, as the fraction of giant planets among the 
multis is smaller than it is for the singles \citep{Latham:2011}, the probability of a multi being a system of 
larger planets orbiting a fainter star is less than that of a single candidate orbiting a star fainter than the 
target.

A second class of multiple FPs can be produced from a single physical 
system, either a triple star with two sets of eclipses or a background 
star with multiple transiting planets.  Triple star systems are unstable 
for small period ratios and produce large eclipse timing variations for 
moderate period ratios, so if large timing variations are not detected, we do not 
need to consider triple star system FPs (nor grazing eclipsing binary stars with a transiting planet) 
unless period ratios are large.  

Quadruple star systems consisting of two pairs of eclipsing binaries \citep[e.g.,][]{Graczyk:2011} could produce a small excess 
probability of a pair of FPs.  Such systems, as well as line of sight pairs of physically-unrelated EBs, also 
might be more likely to mimic planetary-depth transits because of mutual dilution of the fraction of light in 
the \ik aperture.

We note also that FPs are inherently more likely among both small (Earth-size and smaller) and large (Jupiter-size and larger)  candidates than among intermediate (Neptune-size) candidates, independent 
of multiplicity.  Most small candidates produce low-amplitude transits that reduce our ability to remove 
false positives from the candidate list on the basis of the shape of the lightcurve or from differences in the 
centroid of the location of the dip relative to that of the target.  Small candidates, including multis, are 
more likely to be planets orbiting stars fainter than the target simply because the Neptune-size planets 
that could produce such signals are far more common according to \ik observations than are giant 
planets \citep{Borucki:2011}, and this difference in abundance is even larger for the multis than for the 
singles  \citep{Latham:2011}.  At the other extreme of the planet size distribution, giant candidates 
appear to be more contaminated by stellar false positives \citep{Demory:2011} of the type which  plague 
ground-based surveys than do Neptune-size candidates.

(ii): Planets and FPs of both types are more easily detectable for quieter targets; planets and PTs are 
more detectable for brighter targets, but BGEBs that are bright enough to mimic observable transits of the 
target star are more likely for fainter targets. A grazing eclipsing binary with a circumstellar or circumbinary 
transiting planet analogous to Kepler-16 \citep{Doyle:2011} would violate our assumption of a lack of 
correlation between planets and FPs, but such a configuration requires a large period ratio for stability 
and an even larger period ratio for the planet to have modest TTVs.

The 
assumption of no correlation between planets and FPs would be violated if the orientations of orbital planes of PTs with one star having a planet are not random.  Alignments of planetary orbital planes for binary stars for which both stars possess planets also violate this assumption, although members of this 
class are true planets that are all physically associated (albeit less directly) with the target star and thus 
would for many purposes not be classified as FPs.   

The numbers quoted  in Equations (2--9) do not incorporate the caveats noted in this subsection.  
Correlations between the propensity of a target to have one or more FPs and to host one or more 
identified true planets (ii) are likely to be at most a factor of a few. False positives of the BGEB class are 
likely to be concentrated among faint target stars that are located at low galactic latitude 
\citep{Morton:2011}; this concentration increases the expected number of targets with two (or more) FPs, but also 
points to a substantial portion of \ikt's target list whose members have {\it a priori} FP likelihood that is 
lower than or comparable to those given by the numbers above.  

\subsection{Period Ratios: Resonances \& Stability}

The distribution of candidate period ratios exhibits spikes near strong mean motion resonances, which 
implies that candidates with such period ratios are more likely to be true planets than those candidates 
not in or very near such resonances \citep{Lissauer:2011a}. This clustering can be used to increase confidence in candidates with periods near resonances; such a ``near-resonance boost'' can be 
quantified using the measured values of the normalized distance from resonance computed from the distribution of planet period ratios in 
\cite{Lissauer:2011a}. Candidate resonant planets can, however, arise from period aliases in analysis of 
light curves; such aliases are not of the form of the astrophysical FPs considered in this paper, and we 
note that special care to search for them is an important aspect of validating candidate resonant planets 
(see the discussion of KOI-730 and the Note Added in Proof in \cite{Lissauer:2011a}).

Resonant period ratios increase confidence in various candidates, and  as mentioned at the beginning of Section 2.1, one of the 
candidate systems, KOI-284, is highly suspect because two of the candidates have periods that differ 
from one another by $< 4\%$, which would rapidly lead to a dynamical instability on a very short time 
scale (but see the discussion of this system in Section 2.4). Do non-resonant periods reduce the confidence in individual multis and do stable systems 
produce an increase in our confidence?  The answers to both parts of this question are yes, but in most 
cases these effects are minor.  While resonances are overpopulated, a substantial majority of multis are 
not in nor  near such resonances \citep{Lissauer:2011a, Wright:2011}.  And while drawing planetary periods 
randomly from the observed ensemble of transiting planet periods gives many more nominally unstable 
systems than are observed, the fraction of unstable configurations from such a random draw is still small.  
Nonetheless, in high-multiplicity densely-packed systems, wherein most configurations with the same 
number of planets randomly spread over the same range in period would likely be unstable, the boost in 
confidence from stability can be significant. Note that special period ratio boosts and stability boosts are in addition to the standard multiplicity boost, i.e., they can be applied together as two multiplicative factors in the planet prior for appropriate systems.

\subsection{KOI-284 \& Multis Composed of Planets Transiting Differing Stars in a Binary}

A very rough estimate of the expected number of multiple candidate systems that orbit different 
members of a binary/multiple star system (rather than a single system of planets orbiting the same star) 
can be made by assuming that half of \ikt 's targets are binaries, that each star within a binary is as likely 
to host both single and multiple transiting planets as are single stars, and that the probability of one star 
within a binary hosting transiting planets is uncorrelated with that of its binary companion.  These 
assumptions imply $n_c$ planets spread around $3n_t/2$ stars, and thus $(2/3)(n_c/n_t)$ planets per 
star and $2n_c^2/(9n_t) \approx 2$ cases in which multis are composed of planets transiting differing 
members of a stellar binary.  Positive correlations between the propensity of stars in a binary to have 
transiting planets (resulting from the abundance of planets and/or alignment of orbital planes) would 
increase this number, but the fact that a significant majority of binaries are composed of stars of substantially differing 
surface brightness combined with the paucity of giant planets transiting low luminosity stars should reduce 
the number of multis comprised of two planetary systems because of the difficulty of detecting planets 
around secondary stars.

As mentioned above, KOI-284, with three planetary candidates having periods of 6.18, 6.42, 18.0 days 
and nominal sizes of $\sim$ 2 R$_\oplus$ (Earth radii), is the only one of 170 multi-candidate system identified in \cite{Borucki:2011} 
that would be clearly dynamically unstable if all of the planets orbited the same star 
\citep{Lissauer:2011a}\footnote{The other multi that was unstable according to the integrations reported in \cite{Lissauer:2011a} was the 4 candidate system KOI-191. This instability resulted from the strong perturbations of giant planet candidate KOI-191.01 on the nearby outermost candidate KOI-191.04. As the period ratio of these two candidates was 1.258,  \cite{Lissauer:2011a} speculated that the system was librating in a 5:4 mean motion resonance and that this resonance protected the planets from close approaches.  Subsequent data analysis shows that the period of KOI-191.04 was underestimated by a factor of two.  The updated value of the period of this candidate is 38.6516 $\pm$ 0.0012 days. The revised period places the planets much farther apart, and the system is stable with the updated parameters.}.  Analysis of a McDonald observatory spectrum of this bright (\ik magnitude $Kp  = 11.8$) target yields 
a temperature of $\sim$ 6250 K and log$g = 4.5$, which suggest a F7V star that is twice as luminous as 
the Sun.  Speckle images of KOI-284 obtained on 24 June 2010 revealed two stars differing in 
brightness by less than one magnitude and separated from one another by slightly less than 1$''$ 
\citep{Howell:2011}.  Spectroscopic observations of each individual star obtained at Keck in 2011 show 
the two stars to have nearly identical spectra and have a difference in radial velocity of 0.94 $\pm$ 0.1 
km/s.
The nearly identical velocities are consistent with their being gravitationally bound.
Their separation of $0.9''$ at a distance of roughly 500 pc implies a projected separation of 450 AU; the 
relative orbital velocities for two 1 $M_\odot$ stars on a circular orbit with semimajor axis $a$ = 450 AU 
is 2 km/s.  These observations suggest that the three candidates may well all be transiting planets in the 
same stellar system, with one of the stars hosting one of the six day period planets and the other two 
planets orbiting the other star in the binary. Thus, this system would not call into question the statistical 
arguments presented in Section 2.1.  A detailed analysis of both \ik and ground-based data to test this 
hypothesis and elucidate particulars is currently underway \citep{Bryson:2012}.

\section{ Transit Characteristics: Planets vs.~False Positives }

The numbers derived in Section 2 suggest that the sample of multi-planet candidates identified by \cite
{Borucki:2011} is likely to include only a small fraction of false positives, of order 1\%.  But because of 
the uncertainties associated with the caveats discussed in Section 2, as well as any factors that we may 
have omitted, we do not consider it appropriate to view all of \ikt's multi-planet candidates to be validated 
planets at the 99\% level.  
Nonetheless, it is possible to identify candidates with various favorable characteristics as detailed 
below, and to examine the available data for systems individually for signs of potential inconsistencies 
with the planet hypothesis.  See \cite{Haswell:2011} for a more detailed discussion of transit characteristics.

The shape and duration of a candidate transit can be used to distinguish events produced by planets 
from false positives produced by known astrophysical sources. The {\tt BLENDER} code 
\citep{Torres:2011} compares the observed lightcurve to synthetic lightcurves of both a planet transiting the target star and a 
vast array of possible astrophysical false positives involving binary stars and larger planets transiting 
fainter stars. If the planet hypothesis does not provide an acceptable fit to the data, the candidate is not 
viable; candidates for which the planet fit is marginal are called into question, and such candidates are 
not considered further herein. Astrophysical false positives are considered credible if they provide an 
acceptable fit (3$\sigma$) to the data.  Then the {\it a priori} likelihoods of the planet hypothesis and that 
of a false positive are compared, and a probability that the apparent transit is caused by a blend is 
computed.  The process can be very involved, especially for candidates with orbital periods long enough 
that astrophysical false positives could plausibly be produced by objects with eccentric (not tidally-damped) orbits \citep{Fressin:2011, Lissauer:2011b}. 

True planetary transits of the star that is the dominant source of light in the aperture are wavelength-independent (neglecting the small effects 
of stellar limb-darkening and contamination by flux from other stars in the aperture), whereas false positives are not (except in 
the unlikely case that the effective temperatures of the contributing 
stars are similar).
By providing infrared time series spanning times of transit, the Warm 
Spitzer Mission can assist in the validation of many transiting planet 
systems. Several candidates in multiple systems 
have been observed, and they all exhibit achromatic transit depths, 
consistent with planetary signals \citep{Desert:2012}.

The durations of transit-like signatures can be used to assess the probability of the planetary 
interpretation, especially when more than one planet candidate is present for a given target.  Transit and 
eclipse durations depend on several factors, including the masses and radii of the two objects, the 
impact parameter, $b$, which is defined as the minimum projected separation between the center of the 
smaller body (planet in the case of a planetary transit) and that of the larger body (star) measured in 
units of the radius of the larger body, and the period and eccentricity of the orbit.  Central transits (those 
for which the center of the transiting body passes over the center of the transited body) by a planet on a 
circular orbit have durations that vary as the cube root of the orbital period \citep{Kepler:1619} and 
inversely to the cube root of the stellar density \citep{Seager:2003}; these relationships hold for FPs as 
long as the bright star is substantially larger and more massive than the other body. When the system's mass is dominated by the
primary but the radius of the secondary cannot be ignored, then the
duration, $T_{\rm dur}$, (1$^{st}$ contact -- 4$^{th}$ contact,
which corresponds to the values reported in
\cite{Borucki:2011} and in Table 3 of this paper), of a central transit for a circular orbit satisfies
the following proportionality relationship: 
 \begin{equation}
T_{\rm dur} \propto  \frac{R_\star + R_p}{R_\star}P_{\rm orb}^{1/3} .
\end{equation}
\noindent If the transit duration is measured over the time interval when the center of the planet covers 
the disk of the star, the first factor on the right-hand side of Expression (10) is no longer relevant and the 
proportionality is simply to $P_{\rm orb}^{1/3}$. Note that the orbital period, $P_{\rm orb}$, is very accurately 
measured by \ikt, as is the ratio of the planet's radius, $R_p$, to that of the star, $R_\star$, provided 
dilution by nearby stars is small.
 
 Planetary transits with an impact parameter smaller than half the star's radius have a duration that 
exceeds $3^{1/2}/2 \approx 86.67\%$ that of a central transit. Such transits account for almost half of all 
transiting planets, and over half of the transits observed to a specified SNR threshold. The fraction of 
transits with durations greater than half that of a central transit is $3^{1/2}/2$; again specifying a SNR 
threshold increases the observed fraction. Orbital eccentricities also influence the durations of transits 
and FPs; in most cases, durations are reduced because speeds are highest near periastron \citep
{Kepler:1609}, where geometric considerations imply that transit probabilities are larger.

\section{Validation of the 5-planet system Kepler-33}

There are two primary purposes of this work: demonstrating the very high credibility of the ensemble of 
\ik multi-planet candidates and providing a framework to validate particular systems of candidates as 
planets.  We now turn to a specific system that is both intrinsically interesting because it is the closest 
analog to Kepler-11 observed to date \citep{Lissauer:2011a} and has characteristics very favorable for 
validation by multiplicity.  

KOI-707 (KIC 9458613) is a $Kp = 14$ star for which four planet candidates, ranging in period from 13 to 41 days, were 
announced by  \cite{Borucki:2011}; a fifth candidate planet, with shorter orbital period and smaller size, has subsequently been identified.  The lightcurve of this target is shown in Figure 2.

%\epsscale{1.0}
\begin{figure*}
\includegraphics[scale=0.9,angle=270]{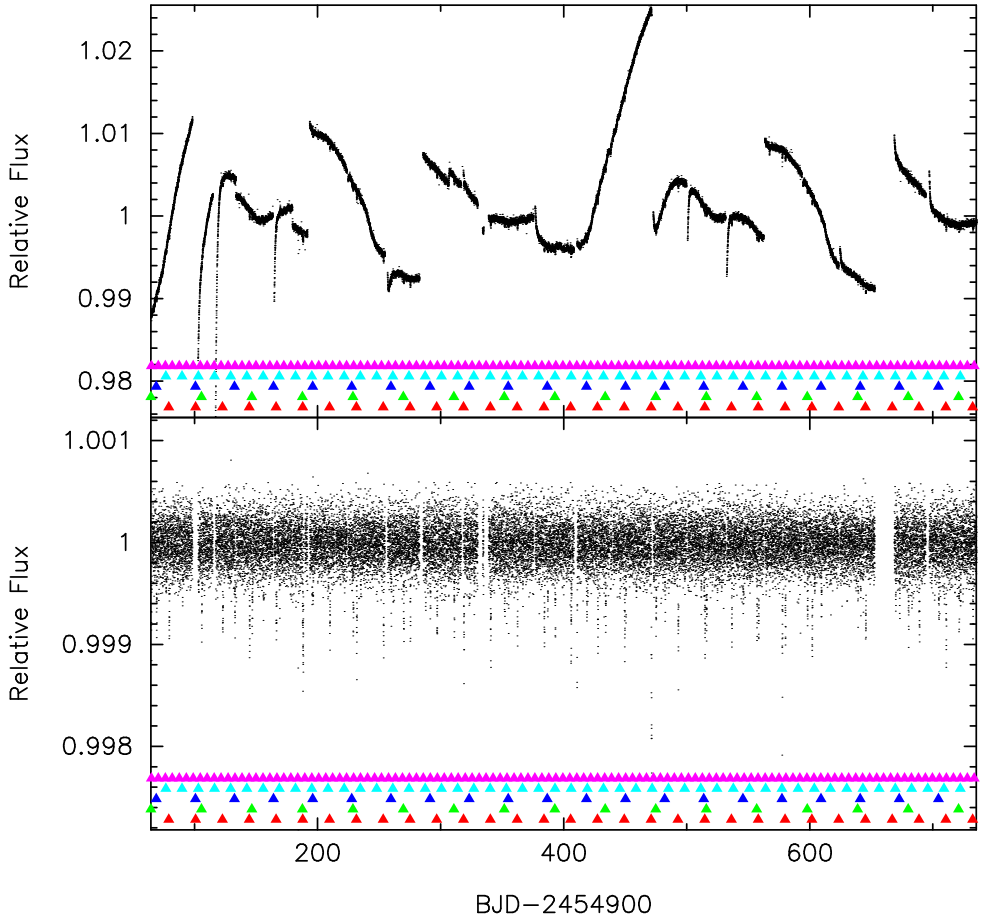}
\caption{\ik photometric time series data for  KOI-707 (KIC 9458613; Kepler-33) in 29.426 minute intervals (standard \ik long cadence observations). Calibrated data 
from the spacecraft with each quarter normalized to its median are shown in the top panel.  The bottom panel displays the lightcurve after detrending with a 
polynomial filter \citep{Rowe:2010}. The entire 8 Quarters of \ik data analyzed herein are displayed.  Note the difference in vertical scales between the two panels.  The 
midpoint times of the five sets of periodic transits (including those that were not observed because they occurred during data gaps) are indicated by triangles of differing colors. Close-ups of folded lightcurves near transit times of each of the planets are shown in Figures 7 and 8. \vspace{0.4 in}}
\label{fig:koi707lightcurve}
\end{figure*}

We performed an analysis of pixel-level \ik spacecraft data that shows 
that the transit signals come from a location on the sky that is coincident with that of the target star within 
small uncertainty, and we obtained high-resolution images of the target that did not show any nearby stars capable of producing the transit signals; these studies are presented in Section 4.1.  Our analyses of the 
transit lightcurves and spectra taken of the target star are presented in Section 4.2.   In Section 4.3, we review the characteristics of this system that allow for its 
validation by multiplicity.  We discuss the properties of the Kepler-33 planetary system in Section 4.4.

\subsection{Location of Transit Signature on the Plane of the Sky}

  A seeing-limited image taken at Lick shows a star $7''$ to the NW that is 4.5 magnitudes fainter than 
KOI-707, but no stars closer to the target.  On 12 June 2011, KOI-707 was imaged in the 880 nm filter ($
\Delta\lambda$ = 50 nm) of our speckle camera at
the WIYN 3.5-m telescope located on Kitt Peak. This star showed no companions within a 2.8 $\times$ 
2.8 arcsec region down to 3.1 magnitudes fainter than the target star; stars within $0.05''$ of the target star 
would not have been detected. Details of this speckle imaging data collection and reduction 
precesses are fully described in \cite{Howell:2011}.

We check to see whether the transit signal is due to a source other than
KOI-707 using the difference image technique described in
\cite{Torres:2011}.  This method fits the measured \ik pixel response
function (PRF) to a difference image formed from the average in-transit
and average out-of-transit pixel images in order to calculate the position of the
difference image centroid.  This difference image centroid position is
compared with the position of the PRF-fit centroid of the average
out-of-transit image, as well as to the predicted deviations in centroid
locations due to scene crowding.  The results for all 5 planet
candidates are presented in Table 1 and 
shown in Figure 3.%\ref{fig:centr1}.

\begin{figure}
\includegraphics[scale=0.27,angle=0]{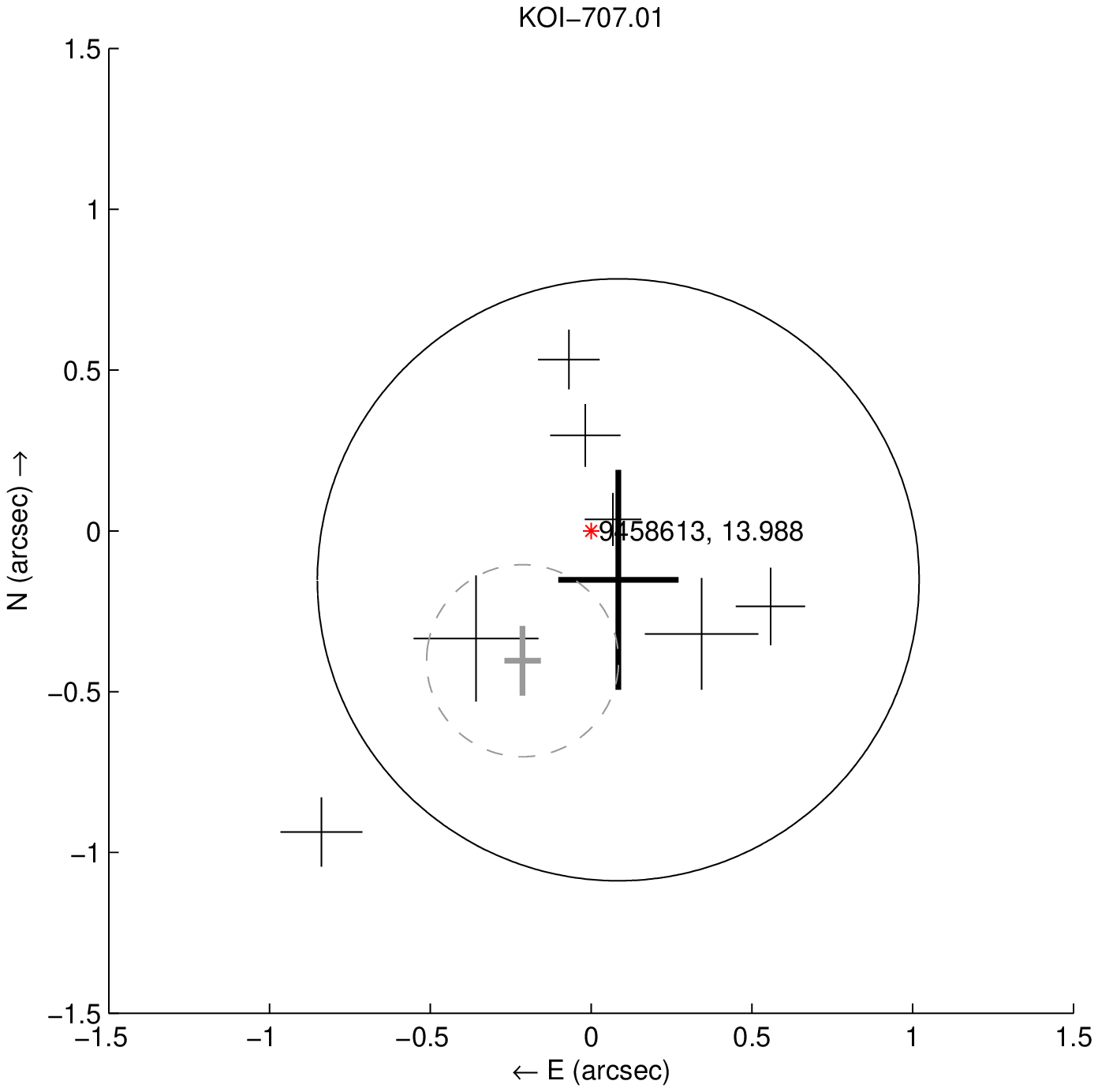}
\includegraphics[scale=0.27,angle=0]{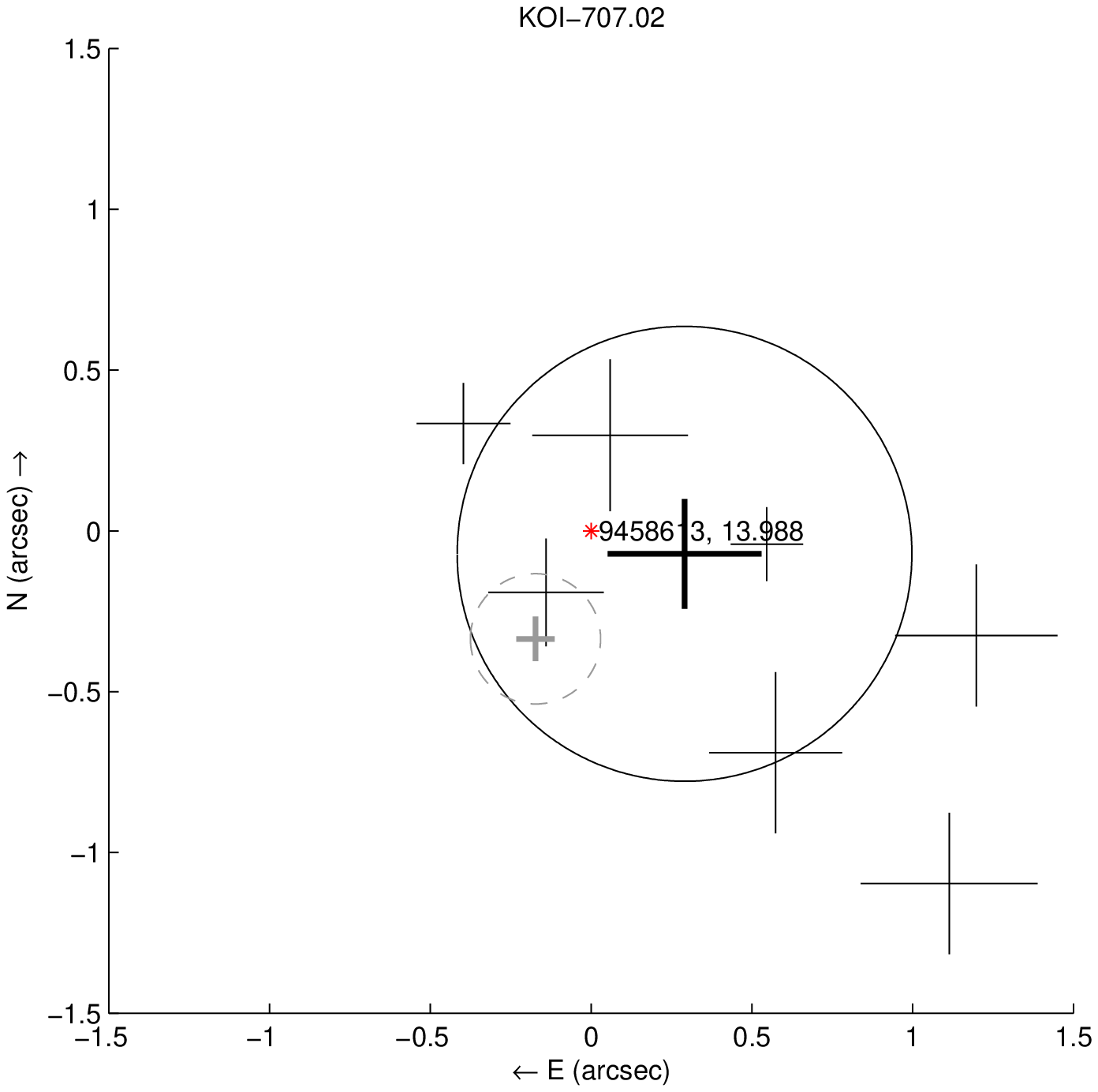}\\
\includegraphics[scale=0.27,angle=0]{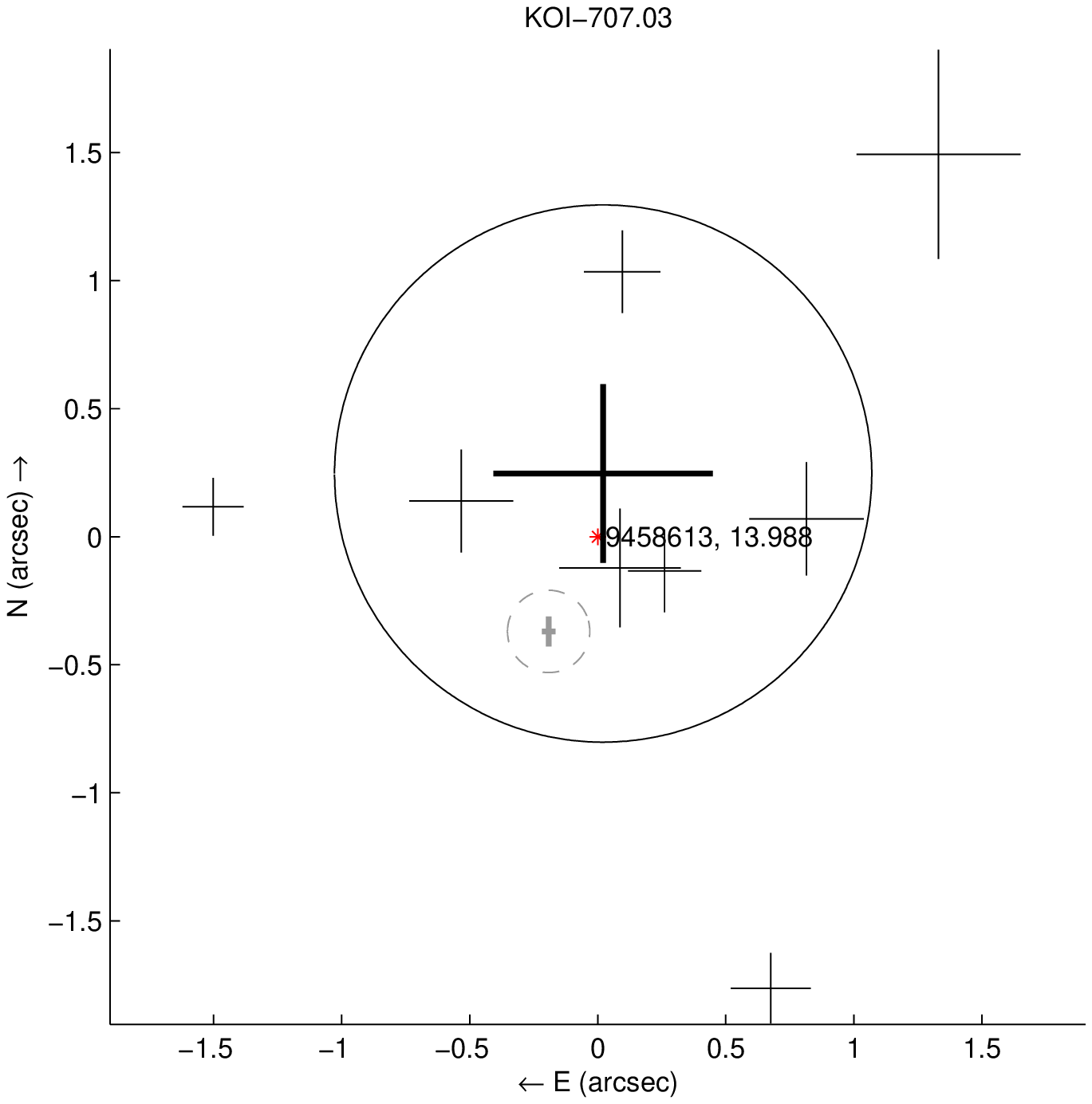}
\includegraphics[scale=0.27,angle=0]{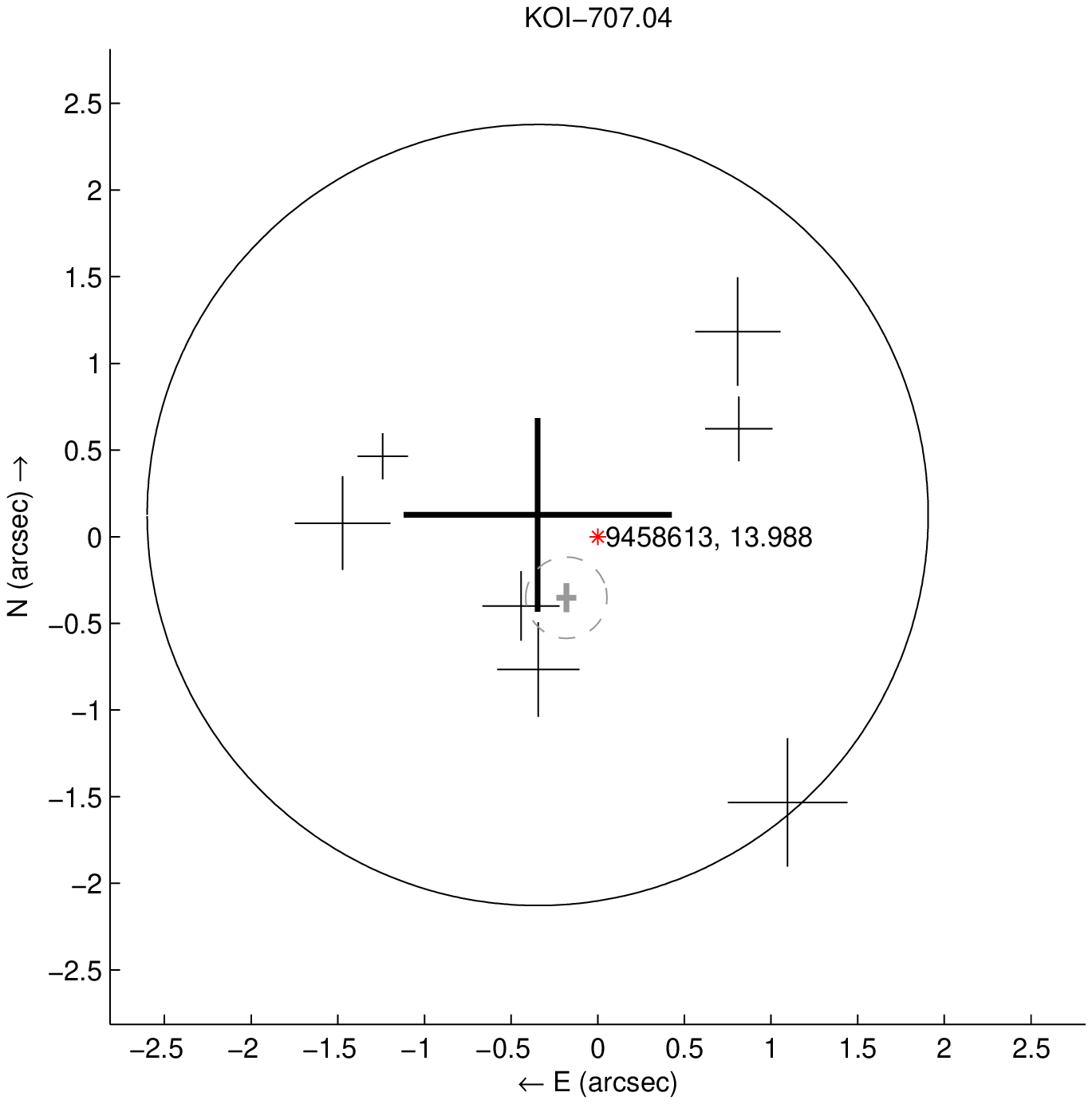}\\
\includegraphics[scale=0.27,angle=0]{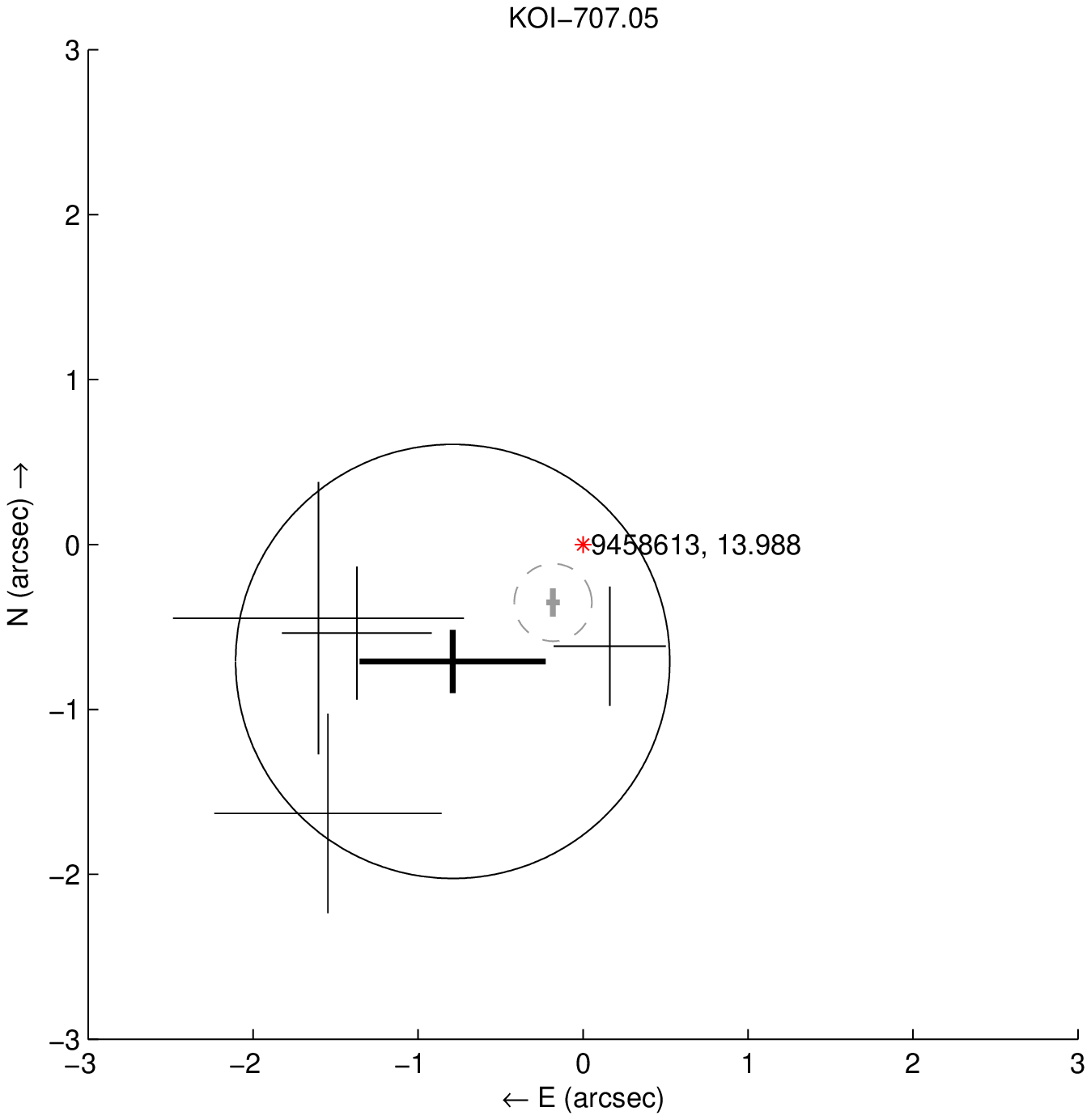}
\caption{Centroid fitting results are shown for all 5 planet candidates
of KOI-707. The origin is the center of light received outside of
transit times.  The red asterisk is the expected centroid of light during transit assuming the transit is on 
the target star; it is displaced from the origin because some of the light entering the aperture comes 
from known background stars whose brightness would not be affected by transits of the target star. (The 
numbers following the red asterisk are the KIC number and \ik magnitude of KOI-707.) The fitted 
centroids of the transit signatures observed
in Quarters 1 through 8 are shown as the thin black crosses, where the arms
of these crosses show the uncertainty in RA and Dec.  Some quarters
as missing for some candidates due to failure of the PRF fit.  The robust
average centroid across quarters is shown by the thick black cross, with the
black solid circle giving the 3$\sigma$ uncertainty in the offset distance.  The robust average
of each modeled quarterly offsets is shown as a thick gray cross, with the gray dashed
circle showing the average model 3$\sigma$ offset distance uncertainty.  When the
circle showing the observed 3$\sigma$ uncertainty in the difference
image has significant intersection with the dashed circle showing
the modeled 3$\sigma$ uncertainty in the position of the target star, the data are consistent with the observed transit signal being on KOI-707. \vspace{0.4 in}
}
\label{fig:centr1}
\end{figure}

The uncertainty in the quarterly centroid locations displayed in Figure 3 %\ref{fig:centr1}  
is based upon propagating pixel-level uncertainty and does not include
possible PRF fit bias.  Sources of PRF fit bias include scene crowding,
because the fit is of a single PRF that assumes a single star, and PRF
error.  The centroid offset is the difference between the positions of the centroids of
the difference image and out-of-transit images, so common biases such as
PRF error should approximately cancel.  There is, however, a quarter-dependent
residual bias.  Statistical analysis across many uncrowded \ik targets indicates that 
the residual bias due to PRF error is zero-mean, so we average the quarterly
centroid positions.  Specifically, we compute a robust 
$\chi^2$ minimizing best-fit position to the quarterly position data.  The uncertainties of the average
are estimated by taking the uncertainties of the fit and performing a bootstrap analysis.
This average for each planet candidate is shown on Figure 3.

Bias due to crowding (stars close enough to
the target that they contribute light observed in the aperture),
however, does not cancel because, to the extent that variations in other
field stars are not correlated with transits, field stars do not
contribute to the difference image.  This introduces a systematic 
bias component that remains in the average over quarters.
To investigate the possibility that
the observed offsets are due to crowding bias, we
model the local scene using stars from the KIC and the measured PRF,
induced the transit on KOI-707 with the ephemeris of each planet
candidate, and performed the same PRF fit analysis on the model
difference and out-of-transit images as that performed on the pixel data.  
If the resulting modeled average offset matches the observed average offset, 
then the observed offset can be explained by crowding bias.

The locations of the transit signatures (difference image centroids) of each of KOI-707's five 
planet candidates lie within $1''$ of the predicted position, and given the uncertainties listed in Table 1,  none of the signatures are significantly
offset and all are fully consistent with transits on the target star. 
Furthermore, the lack of near neighbor stars seen in the speckle image rules out some of the possible 
false positive scenarios that could produce a centroid shift of the magnitude suggested by the marginally 
significant offset observed.

\subsection{Lightcurves, Spectra, Derived Properties}

We obtained a spectrum of the target star KOI-707/Kepler-33 using the HiRES spectrometer at Keck I 10-m telescope. Classification of this spectrum using SME (Spectroscopy Made Easy, \citep{Valenti:1996, Valenti:2005}) yields the stellar parameter values shown in Table 2 (the final two columns in this table show results from constrained fits described at the end of this subsection). The mean stellar density based on spectroscopic information was estimated by matching the determined distributions of $T_{\rm eff}, \log g$ and [Fe/H] to the Yonsei-Yale evolution tracks.  This allowed the determination of stellar mass and radius and mean stellar density posterior distribution based on spectroscopy.  The spectrum of KOI-707 is consistent with two sets of stellar evolution tracks (Figure 4), leading to double-peaked probability distributions and asymmetric uncertainties in several stellar parameters. We plot the probability distributions of the fundamental parameters of stellar mass and age in Figure 5 and those of the stellar density, which is a key input to transit fitting, in Figure 6, and we include in Table 2 various measures of uncertainty beyond the simple standard deviation for the values of various derived stellar parameters that are either critical to the validation and properties of KOI-707 and/or distributed highly asymmetrically.  Note that both solutions are somewhat above the main sequence, implying an evolved star. 

%\epsscale{0.9}
\begin{figure}
\includegraphics[scale=0.63,angle=270]{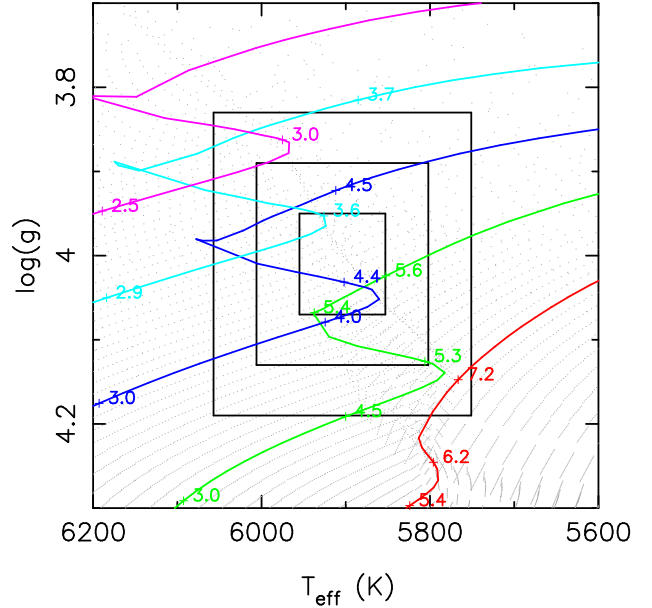}
\caption{Plotted are the Yonsei-Yale evolutionary tracks for 1.1 M$_\odot$ (red), 1.2 M$_\odot$
(green), 1.3 M$_\odot$ (blue), 1.4 M$_\odot$ (cyan) and 1.5 M$_\odot$ (magenta) stars
with [Fe/H] = 0.15.  Ages, in Gyr, are marked along the tracks. Note the uneven rates of evolution along these tracks. The gray dotted tracks are spaced by 0.01  M$_\odot$ in stellar mass and 0.02 Gyr in time to illustrate the distribution of field stars on the $\log g - T_{\rm eff}$ plane; the apparent gaps/discontinuities in these tracks are caused by rapid changes at various stages of stellar evolution. The
three boxes mark the 1, 2 and 3~$\sigma$ confidence regions of the
spectroscopic SME analysis of the spectrum of KOI-707 taken at Keck.   There are
degeneracies in determining a unique stellar mass and radius for specified values of $T_{\rm eff}$ and $\log g$.  For example, a 4.4 Gyr, 1.3 M$_\odot$ model and
a 5.5 Gyr, 1.2 M$_\odot$ model have nearly identical $T_{\rm eff}$ and $\log g$.  The low-mass/old peaks in the probability distribution functions shown in Figure 5 result from the pile-up of stellar evolution tracks as core H is exhausted in the transition regime from radiative cores of 1.1 M$_\odot$ and smaller stars to convective cores in 1.2 M$_\odot$ and larger stars.  The high-mass/young peaks result from the overlap between tracks for the early shell H burning phase of stars with convective cores with tracks for the late core H burning phase of stars that are $\sim 0.12~M_\odot$ more massive. \vspace{0.4 in}
}
\label{fig:evtracks}
\end{figure}

\begin{figure}
\includegraphics[scale=1.6,angle=0]{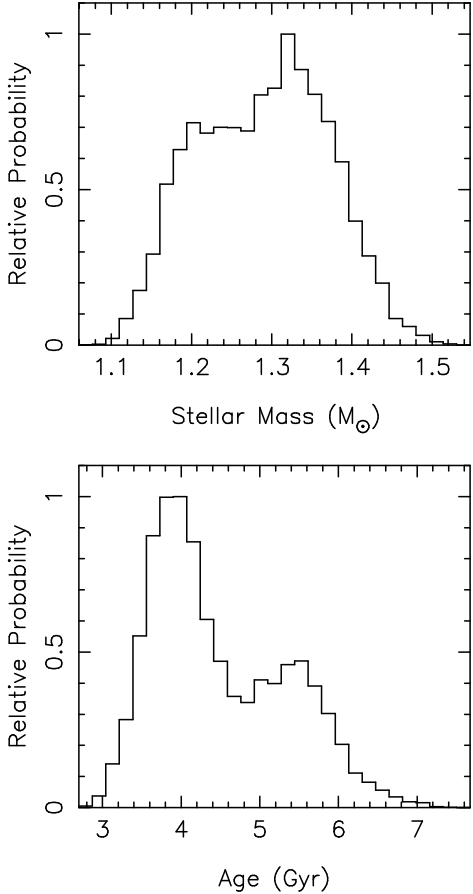}
\caption{ The relative probability distribution of estimates for the mass and age of the star Kepler-33 
(KOI-707), computed using the SME analysis of the star's spectrum and a prior based on 
the frequency distribution of stars, as described in the text. Both curves are double-peaked because two  
solutions, both recently evolved off the main sequence, are consistent with the data.  Integration under 
the curves implies that the younger, more massive, solution is a bit more probable than a somewhat 
older and fainter star.  \vspace{0.4 in}}
\end{figure}

\begin{figure}
\includegraphics[scale=1.6,angle=0]{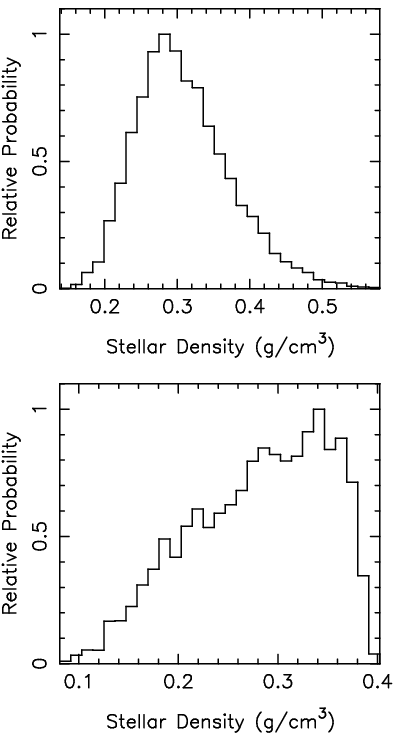}
\caption{ The relative probability distribution of estimates for the density of the star Kepler-33 (KOI-707).  
The top curve was computed using an MCMC analysis of the transit photometry data, whereas the  
results shown in the bottom panel incorporated the SME analysis of the star's spectrum and a prior based on 
the frequency distribution of stars. The spectral solution is highly asymmetric because two groups 
of stars are consistent with the data (see Figure 4).  Note the difference in horizontal scales between the two panels.
\vspace{0.4 in}
}
\label{fig:koi707agemass}
\end{figure}

As an independent check of the values of the SME parameters, we also derived
values by matching the spectrum to synthetic spectra \citep{Torres:2002,
Buchhave:2010}, and in addition we employ a new fitting scheme  (Stellar Parameter Classification, SPC) that
is currently under development and being readied for publication,
allowing us to extract precise stellar parameters from the spectra.
The SPC analysis yielded the following parameters: $T_{\rm eff} = 5849  \pm 50$ K, $\log g = 4.07 \pm 0.10$, [m/H]=$0.12 \pm 0.08, v\sin i= 3.4 \pm 0.5 $ km/s, RV = $13.779 \pm 0.020 $ km/s. The SME and SPC parameters are fully consistent, and the nominal value of the key parameter $\log g$ from SPC is slightly ($< 1 \sigma$) larger than the SME value.  In contrast, the derived stellar radius from SME is $1.82 \pm 0.16~R_\odot$, substantially larger (more than $3\sigma$ formal error, see Table 2) than the radius estimate of 1.29 R$_\odot$ based upon the KIC that was used by used by \cite{Borucki:2011}.         

We measured the radial velocity of KOI-707 relative to the barycenter of our Solar System.  The method involves measuring the Doppler shift of the stellar spectrum obtained with the Keck I telescope and HiRES spectrometer on BJD = 2455782.9441 relative to the solar spectrum by a $\chi^2$ fit of the two spectra.  To set the zero point of the wavelength scale accurately, we measure the displacement of telluric lines from the Earth's atmosphere, serving as the ``iodine cell".   The resulting radial velocities are accurate to 0.1 km/s, based on measurements of 2000 stars \citep{Chubak:2012} and comparison with IAU standard stars.  For KOI-707, the radial velocity is $v_\star$ = 14.09 km/s, indicating motion of the star away from the Solar System at a speed typical of the velocity dispersion of stars in the Galactic disk.
Orbital periods of the planets quoted in Table 3 are as perceived from the barycenter of our Solar System. As the Kepler-33 system is receding from us at $4.7 \times 10^{-5}c$, where $c$ represents the speed of light, the actual orbital periods in the rest frame of the Kepler-33 system are 0.999953 times as large as the tabulated values.

We used Q1-Q8 long-cadence \ik aperture photometry (i.e., only pixel-level corrections such as bias and 
dark smear are applied to the data and there is no co-detrending).  At this level of data 
reduction, there are still instrumental artifacts on timescales similar 
to the planetary orbits.  To detrend the data (remove long-period variations), a second-order polynomial 
was fitted to  segments of sequential \ik photometry data outside of transits (i.e., in-transit observations were given zero weight), and then this polynomial was subtracted from the data.  A 
segment is defined as a length of data that is uninterrupted for  5 or more consecutive long cadences ($\sim$2.5 hours).  Each segment is then combined to 
produce a final time series that is then normalized by the median.

We adopt a simple model to fit the data.  We fit for the mean 
stellar density ($\rho_\star$), and each planet's orbital period ($P_{\rm orb}$), transit epoch ($T_0$), scaled planetary 
radius ($R_p/R_\star$) and impact parameter ($b$).  The planets are 
assumed to be on circular orbits; planet-planet interactions are implicitly accounted for by allowing the 
center of transit time to vary in order to best-fit transit shape  (minimization of scatter on residuals).  The 
transit was 
described with the analytic formulae of \cite{Mandel:2002}, and we adopted a 
non-linear limb darkening law with coefficients fixed based on the \cite{Claret:2011} tabulation for the \ik 
band with stellar parameters from SME 
analysis.  A best fit model was computed using a Levenberg-Marquardt 
algorithm to minimize the $\chi^2$ statistic and initial uncertainties 
were estimated via the construction of a co-variance matrix.

The best fit model and co-variances were used to seed a hybrid-MCMC (Markov Chain Monte-Carlo) 
algorithm to determine posterior distributions of our model parameters. 
  The model is considered hybrid, as we randomly use a Gibbs-sampler and a 
buffer of previous chain parameters to produce proposals to jump to a 
new location in the parameter space.  The addition of the buffer allows 
for a calculation of vectorized jumps that allow for efficient sampling 
of highly correlated parameter space.

A direct comparison of the mean stellar density determined by the 
circular transit-model (0.236 $\pm$ 0.080) and the mean stellar density 
determined from spectroscopy (0.301 $\pm$ 0.067) differ by less than 1$\sigma$.  This shows that we can produce a self-consistent circular model 
in which all five planets orbit the same star and produce stellar parameters 
that agree with spectroscopy.

Close-ups of the lightcurves of each of the planets near planetary transit are shown in Figure 7. Transit durations of all five KOIs vary in proportion to the $P_{\rm orb}^{1/3}$ to within 5\%, and to within 3\%
 when planet sizes are factored in as specified in Expression (10).
The remarkable similarity of these transit durations when appropriately scaled to orbital periods  is apparent in Figure 8, which shows the same data as Figure 7 but with the time coordinate scaled by $P_{\rm orb}^{1/3}$.  Such similarity would be 
the consequence of an ensemble of planets transiting the same star on low eccentricity orbits with similar impact 
parameters.  The impact parameters given in Table 3 are indeed quite similar for all of the planets.  But the inclinations implied to produce these parameters require a very fortuitous arrangement in which tilt to the line of sight is just enough larger in the inner planets than the outer ones.  In contrast, if the system is viewed nearly edge on, i.e., $i \approx 90^\circ$, then both a wider range of impact parameters would be allowed (because transit durations are less sensitive to the precise value of $b$ when $b \ll 1$) and similar values of $b$ imply a very flat system rather than a fortuitous ensemble of tilts. 

\begin{figure}
\includegraphics[scale=1.6,angle=270]{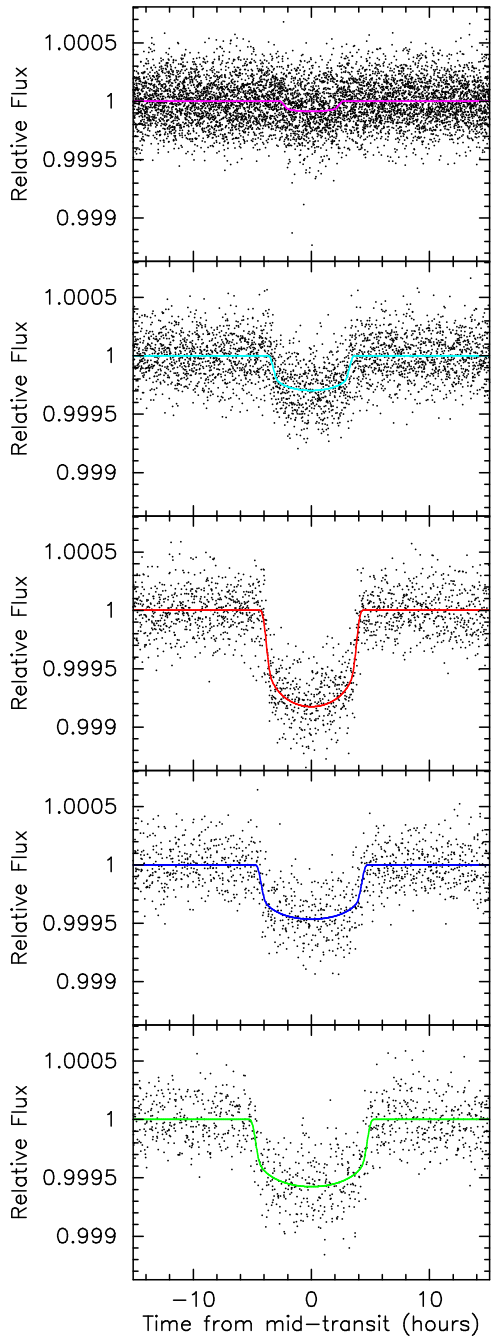}
\caption{Detrended \ik flux from Kepler-33 shown phased at the period of each transit signal and zoomed 
to a 30-hour region around mid-transit. Black dots represent individual \ik long cadence observations.    The folded light 
curves corresponding to the model fits are shown in  colors that
correspond to these used in Figure 2. In each panel, the best-fit model for the other 4 planets was 
removed before plotting.  Each panel has an identical vertical scale, to show the relative depths, and 
identical horizontal scale, to show the relative durations.\vspace{0.4 in}}
\label{fig:koi707transits}
\end{figure}

\begin{figure}
\includegraphics[scale=1.6,angle=270]{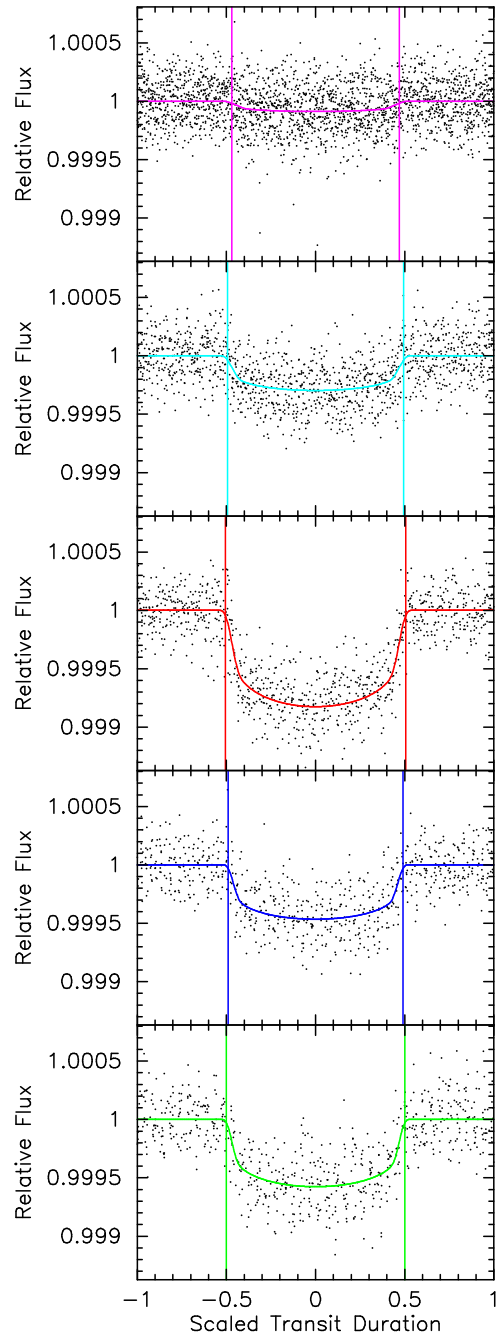}
\caption{The folded lightcurves near the transit of each planet in the Kepler-33 system, as shown in Figure 7, but in this case the time coordinate in each panel has been scaled to  $P_{\rm orb}^{1/3}$ of the planet in question. Vertical lines mark the beginning and end times for each transit. For planets on circular orbits with the same impact parameter, durations normalized in this manner should be the same for all planets.  The remarkable similarity in normalized durations of these planets is very strong evidence that they all orbit the same star.\vspace{0.4 in}  }
\end{figure}

We are thus motivated to perform a second set of fits constrained by the assumptions of circular orbits for all planets and $b = 0$ for the planet with the longest normalized (according to Expression 10) transit duration, KOI-707.01.  Two such fits were performed, one for the low mass stellar solution (where we imposed the constraint $M_\star < 1.21~M_\odot$) and the other for the high mass ($M_\star > 1.21~M_\odot$) solution.  We show in the far right columns of Tables 2 and 3 both the high and low stellar mass ``Flat'' fits; no uncertainties are quoted for these fits because the  $b = 0$ assumption constrains results so tightly that computed uncertainties are unrealistically low.

The fitting procedure that we used to compute the planetary parameters listed in Table 3 did not account for the small (1 -- 2\%) contribution of the second star in the aperture to the flux measured by \ikt.  This dilution implies that the planet radii quoted are overestimated by 0.5 -- 1\%, a difference far smaller than the uncertainties in their actual values. 

\subsection{Validation of the Kepler-33 Planetary System}

The five planetary candidates of KOI-707 are extremely likely to represent a true 5 planet system orbiting 
the target star.  There are several lines of evidence that support this conclusion.  First, the analysis 
presented in Section 2 implies that the overwhelming majority of \ik multiple-planet candidates are true 
planets and bound within the same physical system.  

Second, the planetary system is already closely-packed dynamically for the sizes of planets around an undiluted target star (see Section 4.4). Were the planetary system to orbit another star that is not seen because it is significantly 
fainter than the target, the larger planets required to match the observations would necessitate implausibly low densities for the 
planetary system to be stable.   

Third, the spatial location analysis
presented in Section 4.1, combined with the galactic latitude
distribution of targets and planet candidates shown in Figure 1 and
the relatively high galactic latitude of KOI-707 (15$^\circ$), together imply that
it is unlikely that the signals are due to a source not physically
associated with the target star.  

Fourth, the long durations of the transits relative to the planets' orbital periods imply that they are produced by 
light being blocked from a star whose density is significantly less than those of cool dwarfs, and/or they 
are produced by apocentric transits of objects with substantial orbital eccentricity.  Substantial orbital 
eccentricities would make the system unstable if the planets orbited the same star.  It is highly 
implausible that two stars in an unresolved stellar binary would be so similarly coeval to have nearly 
equal densities characteristic of a somewhat evolved star.  

Fifth, the remarkable similarity of the transit durations when appropriately scaled to orbital periods (Expression (10) and Figure 8) can be 
understood as the consequence of an ensemble of planets transiting the same star with similar impact 
parameters, but would otherwise be an unlikely coincidence.

The above characteristics of KOI-707.01--05, combined with the overall high validity of \ik multis, imply 
that these candidates have a well above 99\% probability of composing a true multi-planet system, hence 
we consider these candidates to be validated planets and name the system Kepler-33.  

\subsection{Characteristics of the Kepler-33 Planetary System}

Numerical integrations to test stability of the system were performed using masses of the planets from 
the nominal mass-radius relationship based upon planets within our Solar System presented in \cite
{Lissauer:2011a}.  This relationship is given by:
 
\begin{equation} M_p = \Big{(}\frac{R_p}{R_\oplus}\Big{)}^{2.06} M_\oplus ,\label{eq:mr} \end{equation} 

\noindent where R$_\oplus$ and M$_\oplus$ are the
 radius and mass of the Earth, respectively.
In previous work,  \cite{Lissauer:2011a} integrated the KOI-707 system using parameters based on the data presented by \cite{Borucki:2011} and assuming planar initially circular orbits; the model system survived intact for the entire $3 \times 10^8$ years simulated. All of our integrations also assume that the system is planar. For all planets having zero initial eccentricity, all three models of the system listed in Tables 2 and 3 survived intact for the entire $10^7$ years simulated.   

Of the three Kepler-33 planetary system models that we integrated, the one based directly on the spectroscopic stellar parameters had the largest planetary sizes and thus largest estimated planetary masses, with the fractional differences in estimated planetary masses from one model to another being larger than those for stellar mass. Thus, the spectroscopic parameters yielded the largest interplanetary perturbations, and presumably the largest TTV signatures.  We examined distributions of TTV signals over 2 year periods during the first 2 Myr of this simulation.  These theoretical TTVs for the inner planet were smaller than 30 seconds, and those for Kepler-33c less than 90 seconds, both values below current measurement capabilities.  The outer three planets exhibited far larger TTVs, 9 -- 18 minutes for Kepler-33d, 23 -- 38 minutes for Kepler-33e and 23 -- 35 minutes for Kepler-33f, roughly consistent with variations detected.  While this model used relatively high mass estimates, the integration began with circular orbits, which produce smaller TTVs, so the results do not yield a preference between the three models of planetary sizes listed in Table 3.  A detailed analysis of planetary TTVs is beyond the scope of this paper, but we note that it may well soon be possible to derive estimates of the masses of some or all of the outer three planets using observed TTVs.

The properties of the 5 planet Kepler-33 system are analogous to those of the 6 planet Kepler-11 system 
\citep{Lissauer:2011b}.  Both systems include 4 planets with orbital periods between 13 and 47 days that are comparable in radius to, or somewhat smaller in size than, 
Neptune, as well as a smaller planet closer to the star.  Both 
systems are thus very closely-packed in a dynamical sense.  The Kepler-11 system is slightly more 
dynamically close-packed than is the Kepler-33 system, both in the sense that the ratios of the orbital 
periods of neighboring planets are a bit larger for Kepler-33's planets \citep{Lissauer:2011a} and as 
evidenced from smaller TTVs detected in the Kepler-33 system.  

In some aspects, the Kepler-33 system can be viewed as a less extreme cousin of Kepler-11, with 5 
known transiting planets rather than 6 and spacing that not quite as close, yet dynamically much tighter than most systems 
\citep{Lissauer:2011a}.  However, the new system's inner planet, Kepler-33b, is smaller and 
closer to its star than is its counterpart, Kepler-11b.  The high density of Kepler-11b relative to its 
brethren, coupled with its smaller size, indicates a composition richer in heavy elements 
\citep{Lissauer:2011b}.  While the mass and thus the density of Kepler-33b is not known, it is unlikely that a planet this 
small ($1.5 - 1.8~R_\oplus$) and close (0.07 AU) to its bright star (3 L$_\odot$) could have retained a substantial H/He atmosphere.  This lends support to the suggestion by  \cite{Lissauer:2011b} that planets such as 
Kepler-11b may have once had atmospheres of light gases that subsequently escaped as a result of 
stellar irradiation.  The slightly shorter normalized duration of the small inner planet Kepler-33b suggests that this planet has somewhat larger inclination and/or eccentricity than do the outer four planets.  We note that Mercury, the innermost and smallest planet in our Solar System, also has the largest eccentricity and inclination of the 8 planets known to orbit our Sun.

Stability of the Kepler-33 planetary system excludes large eccentricities of the four outer planets, although interplanetary 
perturbations must cause eccentricities to be non-zero.  The remarkable similarity in the normalized transit durations of the five planets orbiting Kepler-33 would be very unlikely if the eccentricities and inclinations of planetary orbits are significant (Figure 9).  Moreover, the consequent similarity in computed impact parameters despite the planets' differing orbital distances 
requires that either the system is viewed near edge-on, with impact parameters $b 
\ll 1$, or a very fortuitous coincidence producing an anticorrelation between inclinations and orbital 
periods.  We view such a coincidence as unlikely, and therefore prefer the planetary parameters closer to those listed in final two columns of
Tables 2 and 3 (the ``Flat'' fits) than to those listed in the second column of Tables 2 and 3.

\begin{figure}
\includegraphics[scale=0.45,angle=0]{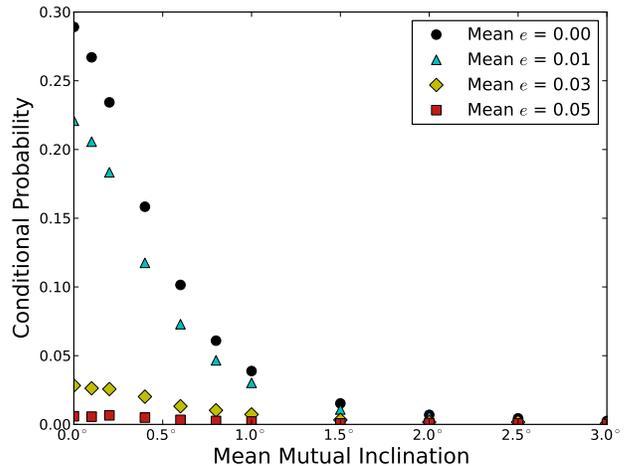}
\caption{The probability of observing all five planets in the Kepler-33 system to have period-normalized transit durations within 3.0\% of one another given that all five planets are
seen to transit the host star is shown  as a
function of mean mutual inclination.  Points were derived using Monte Carlo simulations that assumed parameters from the spectroscopic fit given in the second column of Tables 2 and 3. Results with the flat fit parameters (not shown) are nearly identical.  Mutual inclinations for each realization are drawn
from a Rayleigh distribution with respect to random reference plane
(i.e., random observer). The black dots are for the $e =
0$ case. The blue triangles, yellow diamonds and red squares show results from simulations in which the eccentricity assigned to each
planet was drawn form a Rayleigh distribution with  mean value  $<e>~= 0.01$, $<e>~= 0.03$
and $<e>~= 0.05$, respectively. The addition of even a small range of  eccentricities greatly
reduces probability of highly correlated transit durations. \vspace{0.4 in}}
\end{figure}

\section{Conclusions}

 The calculations presented in Section 2 show that the vast majority of \ikt's multi-planet candidates are 
indeed planets.  The essence of our argument is that fewer than 10$^{-2}$ of \ik targets have a planet 
candidate listed in \cite{Borucki:2011}.  If 10\% or fewer of these candidates are FPs, then fewer than 10$^{-3}$ of \ik targets have a 
false positive, and if FPs are not correlated with other planet candidates, then fewer than 10$^{-5}$ of 
\ikt's $\sim$ 160,000 targets, i.e., a total of at most 2 targets, are expected to have a false positive as well 
as another transit-like signature that makes them a candidate multi-planet system.  This number can be 
compared to 170 candidate multi-planet systems with a total of 408 planet candidates announced by  
\cite{Borucki:2011}. Accurate estimates of the false positive rate in this population require detailed 
analysis of both \ik and ground-based data to account for the caveats expressed in Section 2.2.   
Nonetheless, we expect that $\gtrsim 400$ of the 408 planet candidates in multis that were announced 
by \cite{Borucki:2011} are indeed exoplanets, although a small fraction of these planets orbit stars other 
than their nominal \ik target star.  Note that the smallest planet candidates are most prone to false 
positives, because low SNR limits the restrictiveness in parameter space of tests of lightcurve shape and location of the centroid of the transit 
signal on the sky plane, and the abundance of astrophysical signals that 
are capable of producing lower amplitude events is  greater.  Also note that stellar parameters, and thus planetary sizes, are 
prone to large uncertainties for candidates that have no high-SNR spectroscopic observations from 
the ground. 

From a Bayesian statistical perspective, multiplicity effectively allows us to increase the planet prior, and 
thus enables the validation of candidates as being true planets at a specified high probability level without going to the 
extreme lengths often required to provide a very small upper bound on the FP prior. We use these 
arguments along with a detailed analysis of \ik data to validate the Kepler-11 analog KOI-707 system, which we name Kepler-33.

\acknowledgments 
{\it Kepler} was competitively selected as the tenth Discovery mission. Funding for this mission is 
provided by NASA's Science Mission Directorate. The authors thank the many people who gave so 
generously of their time 
to make the \ik mission a success, chief among them Bill  
Borucki, who has devoted decades to developing and implementing \ikt.  Useful comments were provided by Natalie Batalha, Ruth Murray-Clay, Dimitar Sasselov, Jason Steffen, and especially by Tim Brown.  Kevin Zahnle and Mark Marley provided constructive comments on the manuscript.
D.~C.~F. acknowledges NASA support through Hubble Fellowship grant \#HF-51272.01-A, awarded by STScI, operated by AURA under contract NAS 5-26555.

\begin{deluxetable}{ccccc}
\tablecolumns{5} \tablewidth{0pc} 
\tablecaption{Centroid offsets for KOI-707 based on PRF fits.}
\tablehead{ \colhead{Planet} & \multicolumn{2}{c}{Transit Offset from KOI-707} & \multicolumn{2}{c}{Transit Offset from Prediction} \\
	\colhead{} &
	\colhead{(arcsec)} &
	\colhead{(Offset/$\sigma$)} &
	\colhead{(arcsec)} &
	\colhead{(Offset/$\sigma$)}} 
\startdata
.01 &  $0.17 \pm 0.29$ &  $0.59$ & $0.39 \pm 0.26$ & $1.50$ \\
.02 &  $0.30 \pm 0.22$ &  $1.35$ & $ 0.53 \pm 0.22 $ & $2.38$ \\
.03 &  $0.25 \pm 0.39$  & $0.63$ & $ 0.65 \pm 0.40 $ & $1.64$ \\
.04 & $0.37 \pm 0.93$ & $0.39$ & $ 0.50 \pm 0.67 $ & $0.75$ \\
.05 & $1.06 \pm 0.42$ & $2.53$ & $ 0.71 \pm 0.48 $ & $1.47$ \\
\enddata
\label{tab:centroids}
\end{deluxetable}

\begin{deluxetable}{ccccccccccc}
\tabletypesize{\scriptsize}
\tablewidth{0pc}
\tablecaption{ Characteristics of the Star Kepler-33 \label{tab:star} }
\tablehead{
\colhead{Parameter} & \colhead{Median} & \colhead{$\sigma$} &  \colhead{$+1\sigma$} & \colhead{$-1\sigma$} & \colhead{$+3\sigma$ } & \colhead{$-3\sigma$} & \colhead{Flat, $M_\star>1.2~M_\odot$} & \colhead{Flat, $M_\star<1.2~M_\odot$} }
\startdata
Spectroscopy\\
$M_\star$ (M$_\odot$) &  1.291   &   0.082 &     0.063 &  -0.121 & 0.212 &   -0.200 & 1.271 & 1.164\\
$R_\star$ (R$_\odot$) &  1.82 &   0.16 &     0.14 &  -0.18 &   0.49 &     -0.42 &1.663& 1.615 \\
$Z$ &   0.0250 &   0.0019 &      &   &    &    & 0.0248& 0.0241  \\
$T_{\rm eff} $ (K) &  5904 &   47 &    &  &    & &5899 & 5880 \\
log($L_\star$/L$_\odot$)  &  0.556  &  0.080  &     &    &    & & 0.476 & 0.446   \\
Age (Gyr) &   4.27 &   0.87 &     0.74 &  -1.03 &   2.68 &     -1.65&4.31 & 5.92\\
log$g$ (cm$^2$/s)&   4.027 &   0.056 &     &   &  &   &4.0994   & 4.087\\
$v_\star$ (km/s)&14.09&0.1&&&&&&&\\
$\rho_\star$ (g/cm$^3$) &   0.300 &   0.066 &     0.049 &  -0.079 &   0.230 &     -0.147 &0.3883& 0.3885 \\
Transit Model\\
$\rho_\star$  (g/cm$^3$)&   0.288 &   0.081 &     0.091 &  -0.066 &   0.112 &     -0.215 & 0.03897&0.3897\\
\enddata
\end{deluxetable}

\begin{deluxetable}{ccccccccccc}
\tabletypesize{\scriptsize}
\tablewidth{0pc}
\tablecaption{ Characteristics of the Kepler-33 Planets \label{tab:planets} }
\tablehead{
\colhead{Parameter} & \colhead{Median} & \colhead{$\sigma$} &  \colhead{$+1\sigma$} & \colhead{$-1
\sigma$} & \colhead{$+3\sigma$ } & \colhead{$-3\sigma$} & \colhead{Flat, $M_\star>1.2~M_\odot$} & \colhead{Flat, $M_\star<1.2~M_\odot$} }
\startdata
 Kepler-33b = KOI-707.05\\
$T_0$ (BJD-2454900) & 64.8981 & 0.0075       &      &  & &  & 64.8973 & 64.8973  \\
$P_{\rm orb}$ (days) & 5.66793   &  0.00012  &      &  & &  & 5.667939& 5.667939 \\
$b$ &  0.50   & 0.16   & 0.16     &-0.20  &0.34 &-0.41  &0.298  &0.298 \\
$R_p/R_\star$ &  0.00877 & 0.00046   &      &   &   &    &0.00853&0.00853 \\
$R_p$ (R$_\oplus$) &  1.74&   0.18 &      &   &   &  &1.549 &1.504  \\
$i$ (degrees) &   86.39&   1.17 &     0.96 &  -1.62 &   2.90 &     -3.16 & 88.03& 88.03\\
$a/R_\star $ &  7.87 &   0.87 &   &  &   & & 8.714 & 8.714   \\
$a$ (AU)  &  0.0677  &  0.0014  &      &    &    &&0.06740 & 0.06544    \\
Tr$_{\rm depth}$ (ppm) &   86.8 &   6.8 &     &   &   & & 87.7& 87.7   \\
Tr$_{\rm dur}$ (hr) &   4.88 &   0.16 &     0.16 &  -0.15 &   0.45 &    -0.54 & 4.80& 4.80\\
\\
Kepler-33c = KOI-707.04\\
$T_0$ (BJD-2454900) & 76.6764& 0.0042       &      &  & &  & 76.6777& 76.6777 \\
$P_{\rm orb}$ (days) &  13.17562   &  0.00014  &      &  & &  &  13.17563&  13.17563 \\
$b$ &  0.44   & 0.17   & 0.17     &-0.21  &0.38 &-0.38  & 0.144& 0.144 \\
$R_p/R_\star$ &  0.01602 & 0.00057   &      &   &   &  &0.01560  &0.01560   \\
$R_p$ (R$_\oplus$) &  3.20&   0.30 &      &   &   &   &2.833&2.751  \\
$i$ (degrees) &   88.19&   0.72 &     0.58 &  -1.06 &   1.55 &     -1.92 &89.46&89.46\\
$a/R_\star $ &  13.8 &   1.5 &   &  &   &  &15.292 &15.292   \\
$a$ (AU)  &  0.1189  &  0.0025  &      &    &    & &0.11287  & 0.11484   \\
Tr$_{\rm depth}$ (ppm) &   297.7 &   9.1 &     &   &   &   &299.8  &299.8  \\
Tr$_{\rm dur}$ (hr) &   6.700 &   0.104 &     0.096 &  -0.107 &   0.325 &    -0.295&6.616&6.616 \\ 
\\
Kepler-33d = KOI-707.01 \\
$T_0$ (BJD-2454900) & 122.6342 & 0.0018       &      &  & & &122.6341& 122.6341  \\
$P_{\rm orb}$ (days) &  21.77596   &  0.00011  &      &  & &  &21.775984&  21.7759 \\
$b$ &  0.44   & 0.17   & 0.14     &-0.23  &0.38 &-0.40 & 0 (assumed)  & 0 (assumed)  \\
$R_p/R_\star$ &  0.02667 & 0.00087   &      &   &   &  &  0.02589  &  0.02589   \\
$R_p$ (R$_\oplus$) &  5.35&   0.49 &      &   &   &  & 4.700&4.564   \\
$i$ (degrees) &   88.71&   0.51 &     0.61 &  -0.48 &   1.08 &     -1.37 & 90 (assumed)& 90 (assumed)\\
$a/R_\star $ &  19.3 &   2.1 &   &  &   &  &21.38&21.38   \\
$a$ (AU)  &  0.1662  &  0.0035  &      &    &    &  &0.16533  &  0.16053   \\
Tr$_{\rm depth}$ (ppm) &   831.4 &   10.5 &     &   &   &  &   831&   831   \\
Tr$_{\rm dur}$ (hr) &   8.011 &   0.097 &     0.065 &  -0.103 &   0.345 &    -0.192& 7.987& 7.987 \\
\\
Kepler-33e = KOI-707.03 \\
$T_0$ (BJD-2454900) & 68.8715 & 0.0048       &      &  & & & 68.8720& 68.8720  \\
$P_{\rm orb}$ (days) &  31.78440   &  0.00039  &      &  & & &  31.78438  &  31.78438  \\
$b$ &  0.47   & 0.16   & 0.14     &-0.21  &0.38 &-0.37 & 0.204& 0.204  \\
$R_p/R_\star$ &  0.02011 & 0.00072   &      &   &   &   & 0.01955 & 0.01955  \\
$R_p$ (R$_\oplus$) &  4.02&   0.38 &      &   &   & &3.550  & 3.446  \\
$i$ (degrees) &   88.94&   0.37 &     0.45 &  -0.35 &   0.84 &     -1.07 & 89.576 & 89.576 \\
$a/R_\star $ &  24.9 &   2.8 &   &  &   & & 27.51 & 27.51    \\
$a$ (AU)  &  0.2138  &  0.0045  &      &    &    & &0.2127& 0.20656    \\
Tr$_{\rm depth}$ (ppm) &   467 &   12 &     &   &   &   & 469& 469  \\
Tr$_{\rm dur}$ (hr) &   8.90 &   0.13 &     0.12 &  -0.13 &   0.40 &    -0.38 & 8.819& 8.819 \\
\\
 Kepler-33f = KOI-707.02 \\
$T_0$ (BJD-2454900) & 105.5763 & 0.0040       &      &  & &   & 105.5757 & 105.5757 \\
$P_{\rm orb}$ (days) &  41.02902 &  0.00042  &      &  & & &  41.02903&  41.02903  \\
$b$ &  0.43   & 0.17   & 0.15     &-0.23  &0.39 &-0.42 & 0.130& 0.130  \\
$R_p/R_\star$ &  0.02227 & 0.00076   &      &   &   &  & 0.02171  & 0.02171   \\
$R_p$ (R$_\oplus$) &  4.46&   0.41 &      &   &   & &3.941 & 3.827   \\
$i$ (degrees) &   89.17&   0.34 &     0.33 &  -0.42 &   0.68 &     -0.98 & 89.772& 89.772\\
$a/R_\star $ &  29.5 &   3.3 &   &  &   &   & 32.61& 32.61  \\
$a$ (AU)  &  0.2535  &  0.0054  &      &    &    & &0.2522 & 0.2449    \\
Tr$_{\rm depth}$ (ppm) &   579 &   12 &     &   &   &  & 581 & 581   \\
Tr$_{\rm dur}$ (hr) &   9.87 &   0.13 &     0.11 &  -0.15 &   0.43 &    -0.36 & 9.732 & 9.732\\
\enddata
\end{deluxetable}


\begin{thebibliography}{}

\bibitem[Batalha et al.(2010)]{Batalha:2010} Batalha, N.~M., et al.\ 2010, \apjl, 713, 103 

\bibitem[Batalha et al.(2011)]{Batalha:2011} Batalha, N.~M., et al.\ 2011, \apj, 729, 27 

\bibitem[Borucki et al.(2011)]{Borucki:2011}
Borucki, W. J., et al.~2011,
\apj, 736, 19

\bibitem[Brown et al.(2011)]{Brown:2011}
Brown, T. M., Latham, D. W., Everett, M. R., \& Esquerdo, G. A.~2011,
\aj, 142, 112

\bibitem[Bryson et al.(2012)]{Bryson:2012}
Bryson, S., et al.~2012, in preparation

\bibitem[Buchhave et al.(2010)]{Buchhave:2010}
Buchhave, L. A., et al.~2010, \apj, 720, 1118

\bibitem[Chubak \& Marcy(2012)]{Chubak:2012}
Chubak, C. \& Marcy, G.~W.~2012, in preparation

\bibitem[Claret \& Bloemen(2011)]{Claret:2011}
Claret, A. \& Bloemen, S. \aap, 529, 75

\bibitem[Cochran et al.(2011)]{Cochran:2011}
Cochran, et al.~2011,
\apjs, 197, 7

\bibitem[Demory \& Seager(2011)]{Demory:2011}
Demory, B.-O., \& Seager, S. 2011, 
\apjs, 197, 12

\bibitem[Desert et al.(2012)]{Desert:2012} Desert, J.-M. et al.~2012, in preparation

\bibitem[Doyle et al.(2011)]{Doyle:2011}
Doyle, L. R., et al.~2011,
Science, 333, 1602

\bibitem[Fressin et al.(2011)]{Fressin:2011} Fressen, F., et al.\ 2011, 
\apjs, 197, 5

\bibitem[Graczyk et al.(2011)]{Graczyk:2011} Graczyk, D., et al.\ 2011, arXiv, 1108.0446v1

\bibitem[Haswell(2011)]{Haswell:2011}
Haswell, C.~A. 2011.  Transiting Exoplanets.  Cambridge University Press.

\bibitem[Hayward et al.(2001)]{Hayward:2001} Hayward, T.~L., Brandl, B.,
Pirger, B., Blacken, C., Gull, G.~E., Schoenwald, J., \& Houck, J.~R.\
2001, \pasp, 113, 105

\bibitem[Holman \& Murray(2005)]{Holman:2005}
Holman, M. J., \& Murray, N. W.~2005,
Science, 307, 1288 

\bibitem[Holman et al.(2010)]{Holman:2010}
Holman, M. J., et al.~2010,
Science, 330, 51

\bibitem[Horch  et al.(2011)]{Horch:2011}
Horch, E. P., van Altena, W. F., Howell, S. B., Sherry, W. H., \& Ciardi, D. R.~2011,
AJ, 141, 181

\bibitem[Howard et al.(2011)]{Howard:2011}
Howard, A. W., et al.~2011,
Submitted to \apj\ (arXiv1103.2541)

\bibitem[Howell et al.(2011)]{Howell:2011}
Howell, S. B., Everett, M. E., Sherry, W., Horch, E. \& Ciardi, D. M.~2011,
\aj, 142, 19

\bibitem[Jenkins et al.(2010)]{Jenkins:2010} Jenkins, J.~M., et al.\ 2010, \apjl, 713, L87 

\bibitem[Kepler(1609)]{Kepler:1609} Kepler, J. 1609, Astronomia Nova, Heidelberg

\bibitem[Kepler(1619)]{Kepler:1619} Kepler, J. 1619, Harmony of the World, Linz

\bibitem[Latham et al.(2011)]{Latham:2011} Latham, D.~W., et al.\ 2011, \apjl, 732, 24

\bibitem[Lissauer et al.(2011a)]{Lissauer:2011a}
Lissauer, J. J., et al.~2011a,
\apjs, 197, 8

\bibitem[Lissauer et al.(2011b)]{Lissauer:2011b}
Lissauer, J. J., et al.~2011b,
Nature, 470, 53

\bibitem[Mandel \& Agol(2002)]{Mandel:2002}
Mandel, K., \& Agol, E.~2002
\apj, 579, L171

\bibitem[Morton \& Johnson(2011)]{Morton:2011}
Morton, T. D., \& Johnson, J. A.~2011,
\apj, 738, 170

\bibitem[Ragozzine \& Holman(2010)]{Ragozzine:2010} Ragozzine, D., \& Holman, M.~J.\ 2010, 
submitted to \apj, arXiv:1006.3727 

\bibitem[Robin et al.(2003)]{Robin:2003}
Robin, A.~C.., Reyl{\'e}, C., Derri{\`e}re, S. \& Picaud, S.~2003,
\aap, 409, 523

\bibitem[Robin et al.(2004)]{Robin:2004}
Robin, A.~C.., Reyl{\'e}, C., Derri{\`e}re, S. \& Picaud, S.~2004,
\aap, 416, 157

\bibitem[Rowe et al.(2010)]{Rowe:2010}
Rowe, J.~F., et al.~2010,
\apjl, 713, 150

\bibitem[Seager \& Mall{\'e}n-Ornelas (2003)]{Seager:2003}
Seager, S., \& Mall{\'e}n-Ornelas, G.~2003, 
\apj, 585, 1038

\bibitem[Torres et al.(2002)]{Torres:2002}
Torres, G., Neuhauser, R., \& Guenther, E.~W. 2002, \aj, 123, 1701

\bibitem[Torres et al.(2011)]{Torres:2011}
Torres, G., et al.~2011,
\apj, 727, 24

\bibitem[Troy et al.(2000)]{Troy:2000} Troy, M. et al.\ 2000, \procspie,
4007, 31

\bibitem[Valenti \&  Piskunov(1996)] {Valenti:1996} Valenti, J.~A. \&  Piskunov, N.\ 1996, \aaps, 118, 595

\bibitem[Valenti \& Fischer(2005)]  {Valenti:2005} Valenti, J.~A. \& Fischer, D.~A. 2005, \aaps, 159, 141

\bibitem[Wright et al.(2009)]{Wright:2009} Wright, J.~T., Upadhyay, S., Marcy, G.~W., Fischer, D.~A., 
Ford, E.~B., \& Johnson, J.~A.\ 2009, \apj, 693, 1084 

\bibitem[Wright et al.(2011)]{Wright:2011} Wright, J.~T., et al.\ 2011, \apj, 730, 93 

\end{thebibliography}
\end{document}